\documentstyle{article}

\begin{document}

\def\slash#1{#1 \hskip -0.5em / }

\parindent0pt

\thispagestyle{empty}
\begin{titlepage}
\begin{flushright}
HUB--IEP--94/8 \\
UNIT\"U-THEP-11/94\\
hep-ph/9406220 \\
May 1994
\end{flushright}
\vspace{0.3cm}
\begin{center}
\Large \bf
Effective Meson Lagrangian with Chiral and Heavy Quark
Symmetries from Quark Flavor Dynamics \\
\end{center}
\vspace{0.5cm}
\begin{center}
D.\ Ebert\footnotemark[1],
T.\ Feldmann$^{{\footnotesize{1,2}}}$,  \\
{\sl Institut fuer Physik, Humboldt--Universitaet,\\
 Invalidenstrasse 110, D--10115 Berlin, Germany}
\vspace*{3mm}\\
 R.\ Friedrich$^{{\footnotesize{3,4}}}$,
 H.\ Reinhardt$^{{\footnotesize{4}}}$,  \\
{\sl Institut fuer Theoretische Physik, Universitaet Tuebingen,\\
 Auf der Morgenstelle 14, D--72076 Tuebingen, Germany}
\end{center}
\vspace{0.6cm}
\begin{abstract}
\noindent
By bosonization of an extended NJL model we derive
an effective meson theory which describes
the interplay between chiral symmetry and
heavy quark dynamics.
This effective theory is worked out in the low-energy regime using the
gradient expansion. The resulting effective lagrangian describes
strong and weak interactions of heavy
$B$ and $D$ mesons with pseudoscalar Goldstone
bosons and light vector
and axial--vector mesons. Heavy meson weak de\-cay con\-stants,
coup\-ling con\-stants and the Isgur--Wise func\-tion are predicted
in terms of the model parameters partially fixed from the light
quark sector. Explicit $SU(3)_F$ symmetry breaking effects are
estimated and, if possible, confronted with experiment.
\end{abstract}
\vspace{0.3cm}
\footnotetext[1]{Supported by
 {\it Deutsche Forschungsgemeinschaft} under contract Eb 139/1--1}
\footnotetext[2]{e--mail: feldmann@pha2.physik.hu-berlin.de}
\footnotetext[3]{Supported by a scholarship of the
 {\it Studienstiftung des deutschen Volkes}}
\footnotetext[4]{Supported by
 {\it Deutsche Forschungsgemeinschaft} under contract Re 856/2--2}
\vfill
\end{titlepage}


\setcounter{page}{1}

\section{Introduction}

In the framework of the Standard Model, Quantum Chromodynamics
(QCD) is the theory of strong interaction.
However, due to the complicated nature of QCD, hadrons are usually
described by means of phenomenological effective lagrangians.
It is a big challenge to derive these effective lagrangians describing
the dynamics of strongly interacting light and heavy hadrons
directly from QCD by using suitable non--perturbative
hadronization techniques.
Given the complexity of QCD arising from the self--interactions of
non--Abelian gauge bosons, such a program evidently requires
(possibly crude) approximations on the long way from QCD down to the
effective hadron theory. Simplifications arise, however, since we are
naturally restricted to the low-energy region.
Therefore, from the practical point of view, it is sufficient to
find
an
approximation to QCD which mimics the
essential features of low-energy quark flavor dynamics.
This could be achieved, at least principally, in two
alternative ways: either, by considering quark interactions
mediated
through non--perturbative gluon propagators  or by integrating
out the high--energy components of quarks and gluons.
Clearly, the possible form of such effective quark
lagrangians must be restricted by the underlying
symmetries of QCD, which should be viewed as a guide
to find tractable models of quark flavor dynamics.

In the sector of light quark flavors
$q=(u,d,s)$, QCD possesses an approximate $SU(3)_L \times
SU(3)_R$ chiral symmetry which is spontaneously broken
to $SU(3)_V$, leading to the emergence of (pseu\-do)\-Gold\-stone
bosons $\pi,K,\eta$, which receive their masses by the explicit
breaking of chiral symmetry through current quark masses.
As is well known from current algebra and low energy
theorems, chiral symmetry alone almost entirely
determines the flavor dynamics of light mesons without
need of a detailed knowledge of the underlying gluon
dynamics.
This suggests that a model for quark flavor dynamics
which takes into account spontaneous and explicit
breaking of chiral symmetry could lead to a
realistic effective hadron theory.
Of course, writing down such effective quark lagrangians
in general would include non--renormalizable operators
with dimensionful coupling constants connected with an
intrinsic scale $\Lambda$ above which the model has to
be replaced by the full theory.
In our case this scale is naturally expected to coincide
with the scale of chiral symmetry breaking
$\Lambda_{\chi SB}$, separating the perturbative
region of QCD from the non--perturbative domain.

In the past the Nambu--Jona--Lasionio (NJL) model,
originally formulated for strongly interacting nucleons
\cite{NJL},
has been successfully used to describe the
low--energy light flavor dynamics of QCD.
Bosonization of this model
combined with a gradient expansion leads to an effective meson theory,
which gives a surprisingly realistic description of light
pseudoscalar, vector and axial--vector mesons \cite{Eb82,Eb86}.
(Related work on bosonization of approximate
QCD and effective quark models can also be found in \cite{BOS}.)
The success of the NJL model stems mainly from its global chiral
invariance (for vanishing current masses) which is, however,
spontaneously broken in the ground state.
As a consequence of chiral symmetry the resulting
effective meson lagrangian embodies the
soft--pion theorems, Goldberger--Treiman and KSFR
relations, vector dominance and the integrated
chiral anomaly.
In addition, the bosonized NJL model also provides
a natural explanation of how baryons can emerge as
composite quark--diquark states \cite{r1}
or, in the case of large number of colors $N_c \to \infty$,
as chiral solitons \cite{r2}.
(For a recent
review on these subjects see \cite{Eb94} and references therein.)

While the concept of chiral symmetry and its spontaneous breaking
has been proven extremely successful in understanding the light quark
flavor hadrons, it has to be abandoned for heavy quark flavors which
badly break chiral symmetry.
Recently new important symmetries have
been discovered for heavy quark flavors $Q=b,c,\ldots$
which considerably simplify the description of
heavy--light $(Q\bar{q})$--mesons.
These symmetries, which are not manifest in full QCD, arise
in the limit of infinite heavy quark masses $m_Q \rightarrow \infty$.
This limit $\Lambda \ll m_Q$ is somehow
complementary to the case
$m_q \ll \Lambda$ in the light quark sector.
A systematic expansion of QCD in inverse powers of the
heavy quark mass $1/m_Q$ can be formulated in the framework
of Heavy Quark Effective Theory (HQET) \cite{HQET,IW}.
The relevant degrees of freedom for a heavy quark are then
given by
\begin{equation}
  Q_v(x) = \frac{1+\slash{v}}{2} e^{i m_Q v \cdot x} Q(x)
\quad ,
\end{equation}
where $Q(x)$ is the heavy quark field in the full theory
and $v_\mu$ is the velocity of
the heavy quark with $v^2=1$. The irrelevant (`small' spinor)
degrees of
freedom are integrated out defining an effective lagrangian for the
`large' spinor components $Q_v$.
This effective lagrangian can be expanded into a $1/m_Q$--series
of local
operators, where the leading term reads
\begin{equation}
  {\cal L^{HQET}}=\overline{Q}_v (i v \cdot D) Q_v + O(1/m_Q)
  \label{L-HQET}
  \quad ,
\end{equation}
and $D^\mu$ is the covariant derivative of QCD.
The propagator of the field $Q_v(x)$ in momentum space
is simply
$(v \cdot k+i\epsilon)^{-1}$
where $k_\mu$ describes the residual momentum of the
heavy quark such that its total momentum reads
$p^\mu = m_Q v^\mu + k^\mu$.

The leading term in the
lagrangian (\ref{L-HQET}) manifestly shows two new symmetries.
Spin symmetry is due to the fact, that the coupling of the spin
to the color--magnetic field is as usual a $1/m_Q$--effect
and consequently gluons are blind to the spin of
the heavy quark in the limit $m_Q \to \infty$. This
leads to a mass degeneracy for hadrons
that differ only in the spin of the heavy quark. For example
the masses of $B(5280)$ and $B^*(5325)$ agree within
2\%, those of $D(1870)$ and $D^*(2010)$ within 8\%.
As the leading term in (\ref{L-HQET}) is independent of
the heavy mass $m_Q$, an additional heavy flavor symmetry arises.
Both symmetries relate several form factors of
physical matrix elements between QCD bound states
containing one heavy quark like
$B$, $D$ mesons and  $\Lambda_b$, $\Lambda_c$ baryons.
HQET is then the most powerful tool for determining the parameters
of the standard model in the heavy quark sector, namely the
CKM--matrix elements with heavy quarks, in a model independent
way. Here  fruitful information can be obtained
from weak semileptonic decays of heavy hadrons.

In transitions between two heavy mesons the most drastic
simplification gives rise to a unique form factor,
the Isgur--Wise function $\xi(\omega)$
\cite{IW},
\begin{equation}
 \langle H_{v'}|\overline{Q}_{v'} \Gamma Q_v|H_v\rangle
   = \xi(v\cdot v') {\rm tr}
     \left[ \overline{\cal H}_{v'} \Gamma {\cal H}_v \right]
\quad ,
\label{IW}
\end{equation}
where the ${\cal H}_v$ are the matrix
representations for the heavy mesons containing one
heavy quark with velocity $v$. Also the normalization in the
heavy quark limit is known:  $\xi(v\cdot v'=1)=1$. This is crucial
for determining the CKM--matrix element $V_{cb}$ from
semileptonic $b \to c$ transitions.
The decays of heavy mesons into light mesons are
likewise simplified in the heavy quark mass limit, and one can
derive new relations between form factors \cite{FF}.

HQET is still defined in terms of quark and gluon degrees of
freedom. For practical applications it would, however, be desirable to
reformulate it in terms of hadronic degrees of freedom, which
unfortunately cannot exactly be accomplished.
Stimulated by the success of the NJL model for light quark
flavors, we will extend this model to heavy quark flavors.
One might expect that the NJL model
cannot be applied to heavy quark flavors, for which
the heavy quark mass is larger than the cut--off $\Lambda \approx$
1 GeV, as estimated from the light quark sector.
However, in HQET the heavy quark `on--shell' momentum $m_Q v_\mu$
has already been subtracted such that one is left with the
residual momentum $k_\mu = p_\mu - m_Q v_\mu$.
Indeed, the residual
momentum of the heavy quark in a hadron containing a single heavy
quark arises entirely from its interaction with the light degrees of
freedom (light quark flavors and gluons) and is thus of the same
order as the momenta of the light quarks.
Therefore in an effective quark model
$k_\mu$ should be cut off at the same scale $\Lambda$.
This is the NJL model we are using in the present paper.
We will bosonize
this model in both the light-light and heavy-light sector. The
resulting effective meson theory which describes the interplay between
chiral symmetry and heavy quark symmetry, is studied in a low energy
(gradient) expansion. Thereby we reproduce basically the effective
meson
lagrangians with chiral and heavy quark symmetry introduced previously
on phenomenological grounds \cite{Pha}, where, however, all the
expansion parameters are now microscopically determined by the quark
interaction strength, the quark masses and the cut-off.

The organization of the paper is as follows:
In section 2 we define the extended NJL model with heavy quark flavor
and spin symmetry.
In the subsequent section this model is bosonized.
The resulting effective lagrangian
then describes the interactions of composite light pseudoscalar,
vector and axial--vector mesons with heavy mesons organized
in $(0^-,1^-)$ respectively $(0^+,1^+)$ spin--symmetry doublets.
Including electroweak currents in the generating functional
further determines the weak decay constants of heavy mesons
and the Isgur--Wise function in terms of the model parameters.
Section 4 is devoted to a numerical discussion of our results.
By determining the heavy--light four--quark coupling
constant from heavy meson mass relations
we obtain predictions for weak decay constants
$f_B$, $f_D$ and axial and vector couplings between
light and heavy mesons,
including a detailed estimate of $SU(3)_F$ breaking
effects.
Finally, the slope of the Isgur--Wise function is estimated
and confronted with actual experimental fits.
A short summary and some concluding remarks are given in section 5.
Furthermore some mathematical details are relegated to appendices.


\section{Extended NJL Model }
\label{model}

\subsection{The quark lagrangian}

In the extended NJL model under consideration we add
to the free lagrangian
$${\cal L}_0
   = \overline{q} (i \slash{\partial} - \widehat{m}_0)q
   + \overline{Q}_v (iv\cdot \partial ) Q_v $$
a four--quark interaction--term which is motivated by the
general quark--current structure of QCD\footnote{
Summation over repeated color, flavor and Lorentz indices
is understood.}
\begin{equation}
{\cal L}_{int} = -\frac{\kappa}{2}
 \left(\overline{\psi} \gamma_\mu \frac{\lambda_c^\alpha}{2} \psi
\right) \left(\overline{\psi} \gamma^\mu \frac{\lambda_c^\alpha}{2}
\psi
\right)
\quad .
\label{int}
\end{equation}
Here, light quarks ($q=(u,d,s)^T$) and heavy quarks\footnote{
We will not consider bound states
of a top quark as corresponding life--times are expected to be
too short.}
($Q_v=b \mbox{ or } c$) which are combined in $\psi=({q,Q_v})^T$
are coupled through a universal coupling
constant $\kappa$ of dimension $(\mbox{mass})^{-2}$, and
$\lambda_c^\alpha$ are $SU(N_c)$--color matrices.
Besides a global $SU(N_c)$ symmetry
the lagrangian ${\cal L} = {\cal L}_0 + {\cal L}_{int}$
has the chiral $SU(3)_L \times SU(3)_R$ symmetry
of QCD for vanishing light current masses
$\widehat{m}_0 = {\rm diag} (m_0^u,m_0^d,m_0^s)$.
The symmetries of HQET, $SU(2)_{spin} \times SU(2)_{flavor}$,
are as well included in our model, since the interaction is
independent of the heavy quark mass and spin
(note $\overline{Q}_v \gamma_\mu Q_v = \overline{Q}_v v_\mu Q_v$).

The lagrangian (\ref{int}) separates into a light--light part
(denoted by ${\cal L}_{int}^{ll}$), a heavy--light one
(${\cal L}_{int}^{hl}$) and a heavy--heavy part
${\cal L}_{int}^{hh}$
$${\cal L}_{int} = {\cal L}_{int}^{ll} + {\cal L}_{int}^{hl} +
{\cal L}_{int}^{hh} \quad . $$
In the following ${\cal L}_{int}^{hh}$ is discarded.
The interaction (\ref{int}) acts in the color
$(N_c^2 - 1)$--plet $(\overline{\psi} \psi)$--channel.
For subsequent considerations
it is convenient to Fierz--rearrange this interaction
into the physical relevant (attractive) color--singlet channel.
Defining a coupling constant
$G_1=\kappa (1-1/N_c)/4 $
and
using $SU(3)_F$ matrices $\lambda^a_F$
and $\lambda^0_F=\sqrt{2/3} {\; \bf 1}_F$
with
$ tr_F \left[\lambda_F^a \lambda_F^b \right] =2\delta^{ab}$,
one obtains
\begin{eqnarray}
{\cal L}_{int}^{ll} & =
  &2 G_1 \left(
	  (\overline{q} \frac{\lambda_F^a}{2} q)^2
	 + (\overline{q} i\gamma_5 \frac{\lambda_F^a}{2} q)^2
	\right)
\nonumber\\
&& - G_1
	\left(
      (\overline{q} \gamma^\mu \frac{\lambda_F^a}{2} q)^2
      + (\overline{q} \gamma^\mu \gamma_5 \frac{\lambda_F^a}{2} q)^2
	\right)
\quad ,
\\
{\cal L}_{int}^{hl} &=&
     G_1
        \left(
      (\overline{Q}_vi\gamma_5 q)(\overline{q}i\gamma_5 Q_v)
      - (\overline{Q}_v \gamma^\mu  q) P_{\mu\nu}^\perp
        (\overline{q} \gamma^\nu Q_v)
        \right)
\nonumber\\
  &&   + G_1
        \left(
       (\overline{Q}_v q)(\overline{q} Q_v)
      - (\overline{Q}_v i \gamma^\mu \gamma_5 q) P_{\mu\nu}^\perp
        (\overline{q} i \gamma_5 \gamma^\nu Q_v)
        \right)
\quad ,
\label{lint}
\end{eqnarray}
where additional terms contributing to diquark channels are
subleading in $1/N_c$ and have been discarded.
In order to obtain expression (\ref{lint})
we have decomposed the interaction terms into longitudinal
and transversal parts by means of projection operators
$P_{\mu\nu}^\parallel = v_\mu v_\nu$,
$P_{\mu\nu}^\perp = g_{\mu\nu} - v_\mu v_\nu$,
leading to the following identities
\begin{eqnarray}
(\overline{Q}_v \gamma_\mu q)(\overline{q} \gamma^\mu Q_v) &=&
  (\overline{Q}_v q)(\overline{q} Q_v)
  \nonumber \\ &&
   +(\overline{Q}_v \gamma^\mu  q) P_{\mu\nu}^\perp
  (\overline{q} \gamma^\nu Q_v)
\label{7}
\quad , \\
(\overline{Q}_v i\gamma_\mu\gamma_5 q)
(\overline{q} i\gamma_5 \gamma^\mu Q_v) &=&
  (\overline{Q}_vi\gamma_5 q)(\overline{q}i\gamma_5 Q_v)
  \nonumber \\ &&
   +(\overline{Q}_v i \gamma^\mu \gamma_5 q) P_{\mu\nu}^\perp
  (\overline{q} i \gamma_5 \gamma^\nu Q_v)
\quad .
\label{8}
\end{eqnarray}
Note that the longitudinal components of vector respectively
axial vector currents can be rewritten by means of
$\slash{v} Q_v = Q_v$ in the form of scalar respectively
pseudoscalar currents.
This has been exploited in eqs.\ (\ref{7}), (\ref{8}) to organize
heavy pseudoscalar and vector (respectively
scalar and axial vector) interaction channels
in symmetry doublets of HQET spin symmetry, occuring with the
same interaction strength $G_1$.

\subsection{Generating functional for Greens functions
of \protect\newline quark currents}

The generating functional for Greens functions of quark
bilinears in terms of our model lagrangian is given by
the path integral
\begin{equation}
Z(\eta)= \int {\cal D} \psi {\cal D} \overline{\psi}
e^{i \int d^4x (
{\cal L}_0(\psi) + {\cal L}_{int}^{ll}
  +{\cal L}_{int}^{hl} +
 {\cal L}_{source}(\eta))}
\quad ,
\label{action}
\end{equation}
where we included a term
 ${\cal L}_{source}(\eta)$ containing sources coupled
to weak heavy--light
and  heavy--heavy quark currents of the following type
\begin{eqnarray}
  {\cal L}_{source}(\eta) &=&
  \eta^\dagger (\overline{q} \Gamma Q_v) + h.c.
\label{source1} \\
&& +  \eta_{vv'}^\dagger (\overline{Q}_{v'} \Gamma Q_v) + h.c.
\label{source2}
\quad ,
\end{eqnarray}
where $\Gamma$ is a suitable combination of Dirac matrices.

Expectation values involving the mesonic bound states are
obtained as usual by differentiating with respect to
additionally introduced mesonic sources
and amputating external meson poles.

Following the standard path integral bosonization
procedure \cite{Eb82,Eb86,Eg}, we introduce
color singlet composite $(q\overline{q})$-- and
$(\overline{q}Q_v)$--meson fields in such a way
that the action in (\ref{action}) becomes bilinear
in the quark fields and the latter can be integrated
out.

In the light sector we have
scalar ($ {\rm s}= {\rm s}^a \lambda_F^a/2$),
pseudoscalar (${\rm p} = {\rm p}^a \lambda_F^a/2$),
vector (${\rm v}_\mu ={\rm v}_\mu^a \lambda_F^a/2$) and
axial--vector (${\rm a}_\mu = {\rm a}_\mu^a \lambda_F^a/2$) fields.

In the heavy--light sector (which we will refer
to as heavy in the following) the vector and axial vector fields,
$\phi_\mu$ and $\phi^5_\mu$,
satisfy the constraints
$P_{\mu\nu}^\perp \phi^\nu = \phi_\mu$,
$P_{\mu\nu}^\perp \phi^{5\,\nu} = \phi^5_\mu$,
being equivalent to the transversality condition
$v^\mu \phi_\mu = v^\mu \phi^5_\mu = 0$.
Furthermore, we can collect
the pseudoscalar field $\phi^5$ and the vector field $\phi^\mu$
into a (super)field $h$
which represents the
($0^-,1^-$)--doublet of spin symmetry. Analogously the
scalar field $\phi$ and the axial--vector field $\phi^{5 \, \mu}$
are combined in the parity conjugate (super)field $k$
\begin{eqnarray}
h &=& P_+
      (i \phi^5 \gamma_5 + \phi^\mu \gamma_\mu)
\quad ,\\
k &=& P_+
      (\phi  + i \phi^{5\,\mu} \gamma_\mu \gamma_5)
\quad ,\\
\overline{h} &=& \gamma_0 h^\dagger \gamma_0 =
      (i \phi^{5\,\dagger} \gamma_5 + \phi^{\mu\,\dagger} \gamma_\mu)
P_+
\quad ,\\
\overline{k} &=& \gamma_0 k^\dagger \gamma_0 =
      (\phi^\dagger  + i \phi^{5\,\mu\,\dagger} \gamma_5 \gamma_\mu )
P_+
\quad ,
\end{eqnarray}
where the projection operator on the heavy quark velocity is
defined through $P_+ = (1 + \slash{v})/2$.
This is a shorthand notation, as these fields carry light
flavor quantum
numbers
$h=h^a=(h^u,h^d,h^s),k^a=(k^u,k^d,k^s)$ to form
anti--triplets under chiral symmetry and
the dependence on the heavy quark velocity $h=h_v$, etc.\  has
not been quoted explicitely. Due to flavor symmetry
of HQET these fields describe both $B$ or $D$ mesons.
Note that in case of unbroken chiral $SU(3)_L\times SU(3)_R$
symmetry parity--conjugated
heavy mesons $h$ and $k$ are degenerated.

For later use let us also introduce left and right
combinations
\begin{eqnarray}
  (h+k)_{L,R} &=& (h + k) P_{L,R} \nonumber\\
   &=& P_+ \left( (\phi \mp i\phi^5) + \gamma_\mu
       (\phi^\mu \mp i\phi^{\mu \, 5}) \right) P_{L,R}
\quad ,
\label{lrc}
\end{eqnarray}
where $P_{R,L}= (1 \pm \gamma_5)/2$ are the chiral projectors.

Now we can re--express the generating functional as
\begin{equation}
Z(\eta) = {\cal N} \int {\cal D} \psi {\cal D} \overline{\psi}
{\cal D} M
 e^{i \int d^4x ({\cal L}'
                       + {\cal L}_{source}(\eta))}
\quad ,
\end{equation}
where ${\cal N}$ is an unimportant normalization factor,
${\cal D} M = {\cal D} {\rm s} {\cal D} {\rm p}
 {\cal D} {\rm v} {\cal D} {\rm a}
{\cal D} h {\cal D} k$ stands for the differential of the
several mesonic fields and the
lagrangian is bilinear in the quark fields,
\begin{eqnarray}
{\cal L}' &=& {\cal L}^{ll} + {\cal L}^{hl} \nonumber \quad ,\\
{\cal L}^{ll} & = &
   \overline{q} (i \slash{\partial}
	- ({\rm s} + i\gamma_5 {\rm p})
        + (\slash{\rm v} + \slash{\rm a}\gamma_5) ) q
\nonumber \\
&&-\frac{1}{4G_1} {\rm tr_F}
 \left[ ({\rm s}-\widehat{m}_0)^2 + {\rm p}^2 -2 {\rm v}_\mu
{\rm v}^\mu
				-2  {\rm a}_\mu {\rm a}^\mu\right]
\quad ,\\
{\cal L}^{hl} & = &
   \overline{Q}_v (iv\cdot\partial) Q_v
   -  \overline{Q}_v (h +k) q
   -  \overline{q}   (\overline{h}+\overline{k} ) Q_v
\nonumber \\
&& + \frac{1}{2G_1}
{\rm Tr} \left[ (\bar{h} + \bar{k}) (h-k)\right]
\quad .
\label{eom}
\end{eqnarray}
Here the trace has to be taken over flavor and Dirac indices,
${\rm Tr} = {\rm tr_F tr_D}$.

In the following it will be convenient to use for the light
scalar--pseudoscalar fields the chiral representation
\begin{equation}
{\rm s} + i {\rm p} = \xi_L^\dagger \Sigma \xi_R
\quad ,
\end{equation}
where $\Sigma$ is a hermitian matrix and $\xi_{L,R}$ are
unitary matrices.
The freedom in the choice of $\xi_{L,R}$ reflects
the local hidden symmetry $SU(3)_h$ \cite{Ba}.
Under $SU(3)_L \times SU(3)_R \times SU(3)_h$ the fields
$\xi_{L,R}$, $\Sigma$ transform as
\begin{eqnarray}
&&  \xi_R(x) \to h(x) \xi_R (x) R^\dagger \quad , \qquad
  \xi_L(x) \to h(x) \xi_L (x) L^\dagger \quad ,\\
&& \Sigma \to h(x) \Sigma h^\dagger(x) \quad ,\\
&& L \in SU(3)_L \quad ,\qquad
   R \in SU(3)_R \quad ,\qquad
   h(x) \in SU(3)_h
\quad . \nonumber
\end{eqnarray}
The additional degrees of freedom contained in
$\xi_L$, $\xi_R$ can be gauged away as in
 usual Higgs mechanism, and all our results will be given
later on in unitary gauge
where
\begin{equation}
\xi_R=\xi_L^\dagger=\xi= \exp(i\pi/F)
\label{unit}
\end{equation}
is an element in the coset space $SU(3)_L \times SU(3)_R / SU(3)_V$.
Here $F$ is the bare decay constant, and
$\pi=\pi^a \lambda^a_F/2$ represents the
light octet of (pseu\-do)\-Gold\-stone bosons
associated to spontaneous breakdown of chiral symmetry
through a non--vanishing vacuum expectation value of $\Sigma$.

It is now convenient to define  new `chirally rotated' fields
of constituent quarks $\chi_{L,R}= \xi_{L,R} P_{L,R} \, q $. Under a
chiral rotation of the original quark fields
\begin{equation}
q_R \to R q_R \quad ,\qquad q_L \to L q_L
\end{equation}
the constituent quark fields transform according to the hidden gauge
symmetry
\begin{equation}
\chi_{L,R} \to   h(x) \chi_{L,R}\quad .
\end{equation}
After the chiral rotation the Dirac operator contains the rotated
meson fields
\begin{eqnarray}
 V_\mu \mp A_\mu =
  \xi_{L,R} ({\rm v}_\mu \mp  {\rm a}_\mu
 + i \partial_\mu ) \xi_{L,R}^\dagger
\quad ,\\
(H+K)= (h+k)_{L} \xi^\dagger_{L} + (h+k)_{R} \xi^\dagger_{R}
\quad ,
\label{rot}
\end{eqnarray}
which transform under a left--right transformation according to the
hidden symmetry group
\begin{eqnarray}
 V_\mu \mp A_\mu & \to &  h(x)(V_\mu \mp A_\mu + i\partial_\mu)
 h^\dagger(x)
\quad ,\\
(H+K) & \to & (H+K) h^\dagger(x)
\quad .
\end{eqnarray}
Source terms have to be rotated appropriately. Note that the
light vector field transforms as a gauge field of
hidden local symmetry $SU(3)_{h}$
$$
V^\mu \to h(x) V^\mu h^\dagger(x) + i h(x)\partial^\mu h^\dagger(x) =
h(x) i D^\mu h^\dagger(x)
$$
defining a covariant derivative
\begin{equation}
D^\mu = \partial^\mu - i V^\mu,
\label{cov}
\end{equation}
while the axial--vector field transforms homogeneously
$$ A^\mu \to h(x) A^\mu h^\dagger(x) \quad . $$
The rotated heavy meson fields are still organized in
spin--symmetry doublets
\begin{eqnarray}
H &=& P_+
      (i \Phi^5 \gamma_5 + \Phi^\mu \gamma_\mu)
\quad ,\\
K &=& P_+
      (\Phi  + i \Phi^{5\,\mu} \gamma_\mu \gamma_5)
\quad ,
\end{eqnarray}
where the  $\Phi$'s are related to the original
fields $\phi$ defined by (\ref{lrc}) through (\ref{rot}).

The lagrangian is now expressed in terms of rotated
quark and meson fields
\begin{eqnarray}
{\cal L}^{ll} &=&
   \overline{\chi} (i \slash{\partial}
	- \Sigma
        + \slash{V} + \slash{A}\gamma_5 ) \chi
\nonumber \\
&&-\frac{1}{4G_1} {\rm tr_F}
 \left[ \Sigma^2 -
 \widehat{m}_0 (\xi_L^\dagger \Sigma \xi_R + \xi_R^\dagger \Sigma
 \xi_L)
 \right]
\nonumber\\
&&+\frac{1}{4G_2} {\rm tr_F}
 \left[ (V_\mu - {\cal V}_\mu^\pi)^2
       +(A_\mu - {\cal A}_\mu^\pi)^2
 \right]
\quad ,\\
{\cal L}^{hl} &= &
  \overline{Q}_v (iv\cdot\partial) Q_v
   - \overline{Q}_v (H+K) \chi
   - \overline{\chi}   (\overline{H} +\overline{K} ) Q_v
\nonumber \\
&& + \frac{1}{2G_3}
{\rm Tr} \left[ (\overline{H} + \overline{K})(H-K) \right]
\quad .
\label{gauss}
\end{eqnarray}
Here we have defined the
vector and axial--vector fields induced by the chiral rotation
\begin{eqnarray}
  {\cal V}_\mu^\pi &=&
\frac{i}{2} (\xi_R \partial_\mu \xi_R^\dagger +
             \xi_L\partial_\mu \xi_L^\dagger)
\nonumber \quad ,\\
  {\cal A}_\mu^\pi &=&
\frac{i}{2} (\xi_R \partial_\mu \xi_R^\dagger -
             \xi_L\partial_\mu \xi_L^\dagger)
\nonumber
\end{eqnarray}
which in unitary gauge can be expanded into powers of
the pseudoscalar meson fields $\pi$.
Moreover, following \cite{Eb86}, we have
introduced an independent coupling constant $G_2$ for
the light vector and axial--vector channel, to obtain satisfactory
results for the $\rho$-- and $a_1$--meson masses.
(Note that ${\rm s}$--${\rm p}$--
and ${\rm v}$--${\rm a}$--sectors are separately
invariant under chiral transformations, so chiral symmetry
allows for different coupling constants.)
In the same way chiral and heavy quark symmetries admit an
independent coupling $G_3$ for the heavy meson
sector. As it will turn out, a new coupling $G_3$ is
indeed needed in order to get reasonable predictions for
heavy meson observables.

The generating functional is then given by
\begin{equation}
Z(\eta) = {\cal N} \int {\cal D} \chi {\cal D} \overline{\chi}
{\cal D} Q_v {\cal D} \overline{Q}_v {\cal D} M'
 J(\xi) e^{i \int d^4x ({\cal L}' + {\cal L}_{source}(\eta))}
\quad ,
\label{z}
\end{equation}
where the differential of rotated meson fields in
unitary gauge reads
${\cal D} M' = {\cal D} \Sigma {\cal D} \xi {\cal D} V {\cal D} A
{\cal D} H {\cal D} K$
and $J(\xi)$ is the Jacobian of the chiral rotation,
which gives rise to the integrated chiral anomaly \cite{WZ,Eb86}.

We are now able to integrate out the
quark degrees of freedom. In the first step let us integrate
over
heavy quark fields $Q_v(x)$
 in (\ref{z}).
Since we are neglecting the influence of heavy--heavy mesons,
the resulting heavy quark determinant is trivial and will
be absorbed into the normalization.
Leaving aside source terms which will be separately treated later
when needed for applications,
we get the lagrangian
\begin{eqnarray}
{\cal L}'' & =
& \overline{\chi} i \slash{D} \chi
-\frac{1}{4G_1} {\rm tr_F}
 \left[ \Sigma^2 -
 \widehat{m}_0 (\xi_L^\dagger \Sigma \xi_R + \xi_R^\dagger \Sigma
 \xi_L)
 \right]
\nonumber\\
&&+\frac{1}{4G_2} {\rm tr_F}
 \left[ (V_\mu - {\cal V}_\mu^\pi)^2
       +(A_\mu - {\cal A}_\mu^\pi)^2
 \right]
\\
&& + \frac{1}{2G_3}
 {\rm Tr} \left[ (\overline{H} + \overline{K})(H- K) \right]
\quad ,
\end{eqnarray}
where
\begin{equation}
i \slash{D} = i \slash{\partial}
- \Sigma
        + \slash{V} + \slash{A}\gamma_5
        - (\overline{H}+\overline{K}) (i v \cdot \partial)^{-1}
              (H+ K)
\end{equation}
is the Dirac operator for the light constituent quarks. Here the
last term
represents the effect of the heavy mesons.
Finally, integrating out the light quarks leads to the
quark determinant
\begin{equation}
\det (i \slash{D}) =
\exp ( N_c {\rm Tr} \ln i\slash{D})
\quad .
\label{det}
\end{equation}
To regularize the quark loops arising from
(\ref{det}) we shall use a universal proper--time
cut--off $\Lambda$ which will be fixed from the light meson data.


\section{The Effective Meson Lagrangian}
\label{res}
Expanding the quark determinant (\ref{det}) in powers of the meson
fields leads to the familiar loop expansion given by
Feynman diagrams with
heavy and light mesons as external lines and heavy and
light quarks in internal loops.
Combining the loop expansion with the gradient expansion one finds the
desired effective meson lagrangian.
\subsection{Light sector}

Let us summarize the essential physical results that have
already been obtained for the light sector by
expanding the functional determinant (\ref{det}) in
terms of
light mesonic
fields $\Sigma,V,A$
around their vacuum values.
This has been performed along traditional diagrammatic
quark loop expansion as well as in a heat kernel expansion
\cite{Eb82,Eb86}.

Only the scalar field $\Sigma$ develops a  non--zero vacuum
expectation
value $\langle \Sigma \rangle^i$ that indicates spontaneous breaking
of chiral symmetry.
It has to be identified with the constituent quark mass $m^i$ and is
determined by the Schwinger--Dyson equation
\begin{equation}
 \langle \Sigma \rangle^i = m^i
 = m_0^i +  8 m^i G_1 I_1^i \quad ,
\end{equation}
with $I_1^i$ given in appendix \ref{appC}.
Including explicit flavor symmetry
breaking through different current
quark masses $m_0^u,m_0^d,m_0^s$
triggers the flavor dependence of constituent
quark masses and induced coupling constants and masses of
light mesons. The physical data of light mesons
then fix the parameters of our model $G_1,G_2,\Lambda,\widehat{m}_0$.

We will first give the results in the non--strange sector putting
$m_0^u=m_0^d=m_0$ and afterwards comment on deviations when
considering
strange mesons.
Neglecting quantum fluctuations of $\Sigma$ around
it's VEV, the effective lagrangian  reads \cite{Eb86,Re89}
\begin{eqnarray}
{\cal L}_{light}& = -&\frac{1}{2 g_V^2} {\rm tr_F}
\left[V_{\mu\nu}^2 + A_{\mu\nu}^2\right]
+ \frac{6m^2}{g_V^2} {\rm tr_F} \left[ A_\mu^2 \right] \nonumber \\
&& + \frac{1}{4G_2} {\rm tr_F} \left[ (V_\mu - {\cal V}_\mu^\pi)^2
                             + (A_\mu - {\cal A}_\mu^\pi)^2
				\right]
\nonumber \\
&& + \frac{m_0m}{4G_1} {\rm tr_F}
	\left[\xi_L \xi_R^\dagger + \xi_R \xi_L^\dagger \right]
\quad ,
\label{light}
\end{eqnarray}
with $g_V = \left( 2/3 I_2 \right)^{-1/2}$
and $I_2=I_2^{uu}$ given in appendix \ref{appC} \footnote{In
literature
one often uses the chiral field $U=\xi_L^\dagger \xi_R$.}.

We redefine the fields in order to get the correctly normalized
kinetic terms and remove the mixing between $A^\mu$
and ${\cal A}^{\pi\,\mu}$
\begin{equation}
\widehat{\pi}=\frac{F_\pi}{F} \pi \quad , \qquad
g_V \widehat{V} = V \quad , \qquad
g_V \widehat{A} = A - \frac{M_V^2}{M_A^2} {\cal A}^\pi \quad .
\label{mix}
\end{equation}
These fields are then to be considered as the fields of physical
$\pi,\rho, A_1$ respectively. This yields the following
relations for physical meson masses $M$, the weak pion--decay constant
$F_\pi = 93$ MeV and the  $\rho$--$\pi$--$\pi$--coupling $g_{V\pi\pi}$
\begin{eqnarray}
  M_\pi^2 &=& \frac{m_0 m}{G_1 F_\pi^2} \quad , \qquad
  M_V^2 = \frac{g_V^2}{4G_2} \quad , \qquad
  M_A^2 = M_V^2 + 6m^2  \quad , \nonumber\\
  F_\pi^2 &=&\frac{1}{4G_2}\left(1-\frac{M_V^2}{M_A^2} \right)
  \quad , \qquad
  g_{V\pi\pi}  =  \frac{g_V}{8G_2 F_\pi^2} \quad .
\end{eqnarray}

In addition, we recover the usual KSFR--relations \cite{KSFR}
$$ g_{V\pi\pi} = \frac{a}{2} g_V \quad ,
  \qquad M_V^2 = a F_\pi^2 g_V^2 $$
if the parameter
$$ a = \left(1- \frac{M_V^2}{M_A^2} \right)^{-1} $$ is chosen as
$a=2$.
This choice yields also the Goldberger--Treiman relation
$$ g_\pi = \frac{m}{F_\pi} $$
with $g_\pi = (2 I_2)^{-1/2}$ being the $\pi q \bar{q}$--coupling
constant.

Using $F_\pi = 93$ MeV, $ M_\pi = 140$ MeV, $m_\rho= 770$ MeV,
$ g_{V\pi\pi}= 6$
as input fixes the model parameters $G_1$, $G_2$, $m_0^u=m_0^d$ and
the intrinsic cut--off scale $\Lambda$ for $N_c=3$ to be
\begin{eqnarray}
   G_1 & = & 5.7 \mbox{ GeV}^{-2} \quad ,\nonumber\\
   G_2 & = & 13.8 \mbox{ GeV}^{-2}\quad ,\nonumber\\
   m_0^{u,d} & = & 3\mbox{ MeV}\quad ,\nonumber\\
  \Lambda & = & 1.25 \mbox{ GeV}\quad ,
\end{eqnarray}
together with $m^{u,d}= 300$ MeV.

For strange mesons, all quantities get flavor dependent
$ g_V \to g_{V}^{ij}$, $M_V \to M_V^{ij}, M_A \to M_A^{ij}$,
$F_\pi \to F_\pi^{ij}$, $M_\pi \to M_\pi^{ij}$. For details
we refer the reader to \cite{Eb86}. For our purpose, it is
sufficient to include strange quark effects via
a different constituent quark mass $m_s \approx M_\phi/2 = 510$ MeV,
as well as to introduce different couplings
$ g_V^{us}=
  \left( \frac{1}{6} (I_2^{uu} + I_2^{ss} + 2 I_2^{us})
  \right)^{-1/2}$
and a different decay constant $F_K = F_\pi^{us} \approx 1.2 F_\pi$.

\subsection{Heavy sector}

The on--shell condition for a heavy meson
$\Phi \sim (\bar{q} Q_v)$
is given by
\begin{equation}
  i \partial_\mu \Phi  = \Delta M v_\mu \Phi
\end{equation}
with $\Delta M = M_\Phi - m_Q$ being the mass difference between
the heavy meson (with mass $ M_\Phi$) and the heavy quark.
This is easily seen by substituting
$\Phi' = \exp(-im_Q v\cdot x) \Phi$
in the Klein--Gordon lagrangian for a meson field
$\Phi' \sim (\bar{q} Q)$
\begin{eqnarray}
 {\cal L} &=& \partial^\mu \Phi^{'\dagger} \partial_\mu \Phi'
             - M_\Phi^2 \Phi^{'\dagger} \Phi' \nonumber \\
&=& 2 M_\Phi \left(
   \Phi^\dagger (iv\cdot \partial - \Delta M) \Phi + O(\frac{1}{m_Q})
\right) \quad .
\label{mabs}
\end{eqnarray}
In subsequent applications we will use a normalization, where a
factor $\sqrt{M_\Phi}$ in (\ref{mabs}) is absorbed into
the fields $\Phi$.
In deriving the effective heavy meson lagrangian (\ref{mabs}) from the
quark determinant (\ref{det}), we shall combine the loop expansion
with a low
momentum or gradient expansion.

\subsubsection{The free heavy meson lagrangian}
\label{zh}

\begin{figure}[hbt]
\begin{center}
\unitlength1cm
\begin{picture}(5,3)
\thicklines
\put(0,1.6){\line(1,0){1.75}}
\put(0,1.57){\line(1,0){1.75}}
  \put(0.8,1.8){\mbox{$p$}}
\put(3.25,1.6){\line(1,0){1.75}}
\put(3.25,1.57){\line(1,0){1.75}}
  \put(4,1.8){\mbox{$p$}}
\put(2.5,1.5){\oval(1.5,1.5)[t]}
\put(2.5,1.47){\oval(1.5,1.5)[t]}
  \put(2.4,2.4){\mbox{$k$}}
\thinlines
\put(2.5,1.5){\oval(1.5,1.5)[b]}
  \put(2.1,0.4){\mbox{$k-p$}}
\put(0,1.4){\line(1,0){1.75}}
\put(3.25,1.4){\line(1,0){1.75}}
\put(1.75,1.5){\circle*{0.2}}
\put(3.25,1.5){\circle*{0.2}}
 \put(-1,1.4){\mbox{$H,K$}}
 \put(5.1,1.4){\mbox{$\overline{H},\overline{K}$}}

\end{picture}
\end{center}
\caption{Self--energy diagram for heavy meson fields $H,K$.}
\label{dia1}
\end{figure}
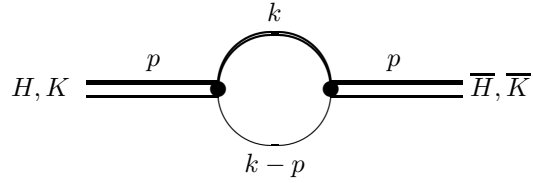

The loop expansion of the fermion determinant (\ref{det}) gives rise
to
the self--energy diagram for the
heavy fields in Figure \ref{dia1}
which yields the following expression
for each single light quark flavor with mass $m^i$
\begin{eqnarray}
&& - {\rm tr_D}
  \left[ \overline{H}^i \Pi_H^i (v\cdot p) H^i \right]
 +  {\rm tr_D}
  \left[ \overline{K}^i \Pi_K^i (v\cdot p) K^i \right]
\nonumber \\
&=&
 i N_c \int^{reg} \frac{d^4k}{(2\pi)^4}
  \frac{ {\rm tr_D} \left[ (\slash{k} -\slash{p} + m^i)
                  (\overline{H}^i + \overline{K}^i) (H^i + K^i)
  \right]}
  {\left( (k-p)^2-(m^i)^2 \right)\left(v\cdot k + i\epsilon\right)}
\quad .
\label{se}
\end{eqnarray}
Expanding the self--energy part
$\Pi_{H,K}^i (v\cdot p)$
in powers of the external momentum $v\cdot p$,
\begin{eqnarray}
\Pi_{H,K}^i(v\cdot p) = \Pi_{H,K}^i(0) +
\Pi_{H,K}^{'\,i}(0) \, v\cdot p
+ O((v\cdot p)^2)
\end{eqnarray}
yields the quadratic meson lagrangian
\begin{eqnarray}
{\cal L}_{heavy} & = &
- {\rm tr_D}
\left[ \overline{H}^i
       \left( - \frac{1}{2G_3}
              + \Pi_H^i (0) + \Pi^{'\, i}_H (0) \, v\cdot p
       \right) H^i \right]
\nonumber \\
&&
+ {\rm tr_D}
\left[ \overline{K}^i
       \left( - \frac{1}{2G_3}
              + \Pi_K^i (0) + \Pi^{'\, i}_K (0) \, v\cdot p
       \right) K^i \right]
\end{eqnarray}
The here obtained lagrangian takes the standard form when we choose
$\Delta M_{H,K}^i$ such that
\begin{eqnarray}
- \frac{1}{2G_3} + \Pi_{H,K}^i(0)
+\Pi^{'\, i}_{H,K}(0) \,  \Delta M_{H,K}^i = 0
\quad
{}.
\end{eqnarray}
Finally rescaling the meson fields by Z--factors,
$Z_{H,K}^i \equiv \left( \Pi^{'\, i}_{H,K} (0) \right)^{-1}$,
\begin{eqnarray}
\widehat{H}^i &=& (Z_H^i)^{-1/2} H^i
\quad , \nonumber \\
\widehat{K}^i &=& (Z_K^i)^{-1/2} K^i
\quad ,
\label{scale}
\end{eqnarray}
the effective meson lagrangian acquires in configuration
space the desired form
\begin{eqnarray}
 {\cal L}_0^{heavy} & = & - {\rm tr_D}
            \left[ \widehat{\overline{H}}^i (iv\cdot \partial - \Delta
	    M_H^i) \widehat{H}^i
            \right]
\nonumber \\
          &&     +  {\rm tr_D}
            \left[ \widehat{\overline{K}}^i (iv\cdot \partial - \Delta
	    M_K^i) \widehat{K}^i
             \right]
\quad .
\end{eqnarray}
The explicit expressions for the $Z_{H,K}^i$ and
$\Delta M_{H,K}^i$ read
\begin{eqnarray}
&&  Z_{H,K}^i  =  \left(I_3^i \pm 2 m^i I_2^{ii} \right)^{-1}
\quad ,
\label{Zfac} \\
&&  \Delta M_{H,K}^i = Z_{H,K}^i
     \left( \frac{1}{2G_3} - I_1^i \mp m^i I_3^i \right)
\quad ,
\label{selfcon}
\end{eqnarray}
where the integrals $ I_1^i, I_2^{ii}$ and $I_3^i$ are given in
appendix \ref{appC}.
Due to heavy flavor symmetry,
mass differences $\Delta M^i_{H,K}$  do not
scale with
the heavy quark mass. In addition, we observe a mass--splitting
between $H$ and $K$ induced through the light constituent
quark mass $m$ which is therefore
an effect of spontaneous chiral symmetry breaking. We will
discuss numerical results following from (\ref{Zfac}),
(\ref{selfcon}) for heavy meson masses
and decay constants later.

\subsubsection{Coupling of heavy mesons $H,K$
to light vector-- and axial--vector--mesons}

\begin{figure}[hbt]
\begin{center}
\unitlength1cm
\begin{picture}(5,3)
\thicklines
\put(0,1.6){\line(1,0){1.75}}
\put(0,1.57){\line(1,0){1.75}}
  \put(0.8,1.8){\mbox{$p_i$}}
\put(3.25,1.6){\line(1,0){1.75}}
\put(3.25,1.57){\line(1,0){1.75}}
  \put(4.1,1.8){\mbox{$p_j$}}
\put(2.5,1.5){\oval(1.5,1.5)[t]}
\put(2.5,1.47){\oval(1.5,1.5)[t]}
  \put(2.4,2.4){\mbox{$k$}}
\thinlines
\put(2.5,1.5){\oval(1.5,1.5)[b]}
  \put(0.9,0.8){\mbox{$k-p_i$}}
\put(0,1.4){\line(1,0){1.75}}
\put(3.25,1.4){\line(1,0){1.75}}
\put(1.75,1.5){\circle*{0.2}}
\put(3.25,1.5){\circle*{0.2}}
\put(2.5,0.75){\circle*{0.2}}
\put(2.4,0.75){\line(0,-1){1}}
\put(2.6,0.75){\line(0,-1){1}}
  \put(3.2,0.8){\mbox{$k-p_j$}}
  \put(2.7,0.1){\mbox{$p_i-p_j$}}
 \put(-1,1.4){\mbox{$H,K$}}
 \put(5.1,1.4){\mbox{$\overline{H},\overline{K}$}}
 \put(2.2,-0.6){\mbox{$V,A$}}
\end{picture}
\end{center}
\caption{Vertex diagram describing
$(\overline{H},\overline{K})$--$(H,K)$--$(V,A)$ couplings.}
\label{dia2}
\end{figure}
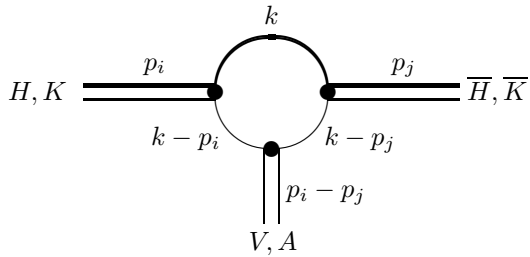

For applications in heavy--flavor decay processes it is necessary
to know the strong interaction couplings
between heavy mesons and light mesons $V$, $A$.
The loop expansion of the quark determinant (\ref{det}) yields such
vertex terms through
the following expression
\begin{eqnarray*}
&&  -\frac{iN_c}{16\pi^4} \int^{reg}
 \frac{d^4k}{((k-p_i)^2 - (m^i)^2)((k-p_j)^2 - (m^j)^2)
    (v\cdot k + i\epsilon)}
\nonumber \\
&& \times {\rm tr_D} \left[ (\slash{k} - \slash{p}_i + m^i)
                         (\slash{V}^{ij} + \slash{A}^{ij} \gamma_5)
                         (\slash{k} - \slash{p}_j + m^j)
                         (\overline{H}^j + \overline{K}^j)
                         (H^i + K^i) \right]
\end{eqnarray*}
which corresponds to the diagram shown in Figure \ref{dia2}.

In the low--momentum expansion
around $v\cdot p_i = v \cdot p_j = 0$
we get the following
contributions to an effective lagrangian in terms
of renormalized heavy and light fields
\begin{eqnarray}
{\cal L}_{V/A}^{heavy} =
   g_V^{ij} \lambda^{ij}_1
 {\rm tr_D}
     \left[ \widehat{\overline{H}}^j \widehat{H}^i \right]
v \cdot \widehat{V}^{ij}
&& - g_V^{ij} \lambda^{ij}_2
 {\rm tr_D}
     \left[ \widehat{\overline{K}}^j \widehat{K}^i \right]
v \cdot \widehat{V}^{ij}
\nonumber\\
   + g_V^{ij} \lambda^{ij}_3
 {\rm tr_D}
     \left[ \widehat{\overline{H}}^j \widehat{H}^i
            \slash{\widehat{A}}^{ij} \gamma_5 \right]
&& + g_V^{ij} \lambda^{ij}_4
 {\rm tr_D}
     \left[ \widehat{\overline{K}}^j \widehat{K}^i
            \slash{\widehat{A}}^{ij} \gamma_5 \right]
\nonumber\\
   - g_V^{ij} \lambda^{ij}_5
 {\rm tr_D}
     \left[ \widehat{\overline{K}}^j \widehat{H}^i
            \slash{\widehat{V}}^{ij} \right]
&& + h.c.
\nonumber\\
   + g_V^{ij} \lambda^{ij}_6
 {\rm tr_D}
     \left[ \widehat{\overline{K}}^j \widehat{H}^i
            \slash{\widehat{A}}^{ij} \gamma_5 \right]
&&+ h.c.
\label{lambda}
\end{eqnarray}
The expressions for the several coupling parameters
$\lambda_n^{ij}$ are
given in appendix \ref{appD}.

In the flavor symmetry limit $m_i=m_j$ due to
$\lambda_1^{ii} = \lambda_2^{ii} = 1$ and $g_V^{ij}=g_V$,
we observe that
the light vector mesons couple via a covariant derivative
of hidden symmetry
\begin{equation}
 D_\mu^{ij} = \partial_\mu \delta^{ij} + i g_V \widehat{V}_\mu^{ji}
\quad .
\end{equation}
(Note the different sign compared to (\ref{cov})
which is due to the fact that the heavy fields $H,K$
transform as anti--triplets under $SU(3)_h$.)

\subsubsection{Coupling of heavy mesons $H,K$
to  pseudo--Goldstone $\pi$'s}

In our approach, the heavy mesons $H,K$ couple to
the fields
$\pi$ via $\pi -A_1$ mixing due to the coupling between $A^\mu$ and
$\cal A^{\pi\,\mu}$ (see eq. (\ref{mix})).
For concreteness, we consider the coupling between two members
of the ($0^-,1^-$)--multiplet
$\widehat{H}^i$ with a pseudoscalar $\widehat{\pi}^{ij}$
\begin{eqnarray}
 {\cal L}_{\pi}^{heavy} =   g_{HHA}^{ij} {\rm tr_D}
     \left[ \widehat{\overline{H}}^j \widehat{H}^i
      \slash{{\cal A}}^{\hat{\pi}\,ij} \gamma_5 \right] + \ldots
\end{eqnarray}
where the coupling is then  given by
\begin{equation}
g_{HHA}^{ij} = \frac{(M_V^{ij})^2}{(M_A^{ij})^2} \lambda_3^{ij}
\quad .
\label{gHHA}
\end{equation}
The ellipsis denote analogous couplings
of $H$ and $K$ where one simply has to
replace $\lambda_3$ by $\lambda_4$ or $\lambda_6$ respectively,
(cf. eq.(\ref{lambda}))
\begin{equation}
g_{KKA}^{ij} = \frac{(M_V^{ij})^2}{(M_A^{ij})^2} \lambda_4^{ij}
               \quad ,\qquad
g_{HKA}^{ij} = \frac{(M_V^{ij})^2}{(M_A^{ij})^2} \lambda_6^{ij}
{}.
\end{equation}
These terms describe direct processes
with an odd number of
Goldstone bosons. The corresponding decays into an even number
of $\pi$'s are given by coupling the vector mesons in (\ref{lambda})
to the pion current
${\cal V}^{\pi}$ in (\ref{light}).

\subsubsection{Electroweak decays of heavy mesons $H,K$}

Next, we present the results for
the electroweak decay constant of heavy mesons $f_{H,K}$ which
determines
the matrix elements of electroweak heavy--to--light
currents between a heavy meson state
and an arbitrary number of Goldstone--fields.
For this purpose we
introduce a source term according to (\ref{source1})
\begin{equation}
  \eta^{\mu\dagger}_L (\overline{\chi} \xi_L
  \gamma_\mu (1-\gamma_5){Q}_v)
  + h.c.
\label{current}
\end{equation}
The source can be removed from the Dirac operator by a simple
shift in the fields $\Phi$ and $\Phi^\mu$ and appears afterwards in
the quadratic term in (\ref{gauss})
\begin{eqnarray}
&&  \frac{1}{2G_3} {\rm Tr} \left[
  (\overline{H}+\overline{K}) (H-K) \right]
\to \nonumber \\
&& \frac{1}{2G_3} {\rm Tr}
       \left[ (\overline{H} + \overline{K}
               + \eta_L^{\mu\dagger} \xi_L \gamma_\mu (1 - \gamma_5))
 (H - K + \eta_L^{\mu}
          \gamma_\mu (1 + \gamma_5)
          \xi_L^\dagger ) \right]
 .
\end{eqnarray}
By variation with respect
to $\eta_L^{\mu\dagger}$ and setting sources equal
to zero afterwards, we
get the following expression for the bosonized current in terms
of the rescaled fields of eqs.
(\ref{scale})
\begin{eqnarray}
&& \frac{1}{2G_3} {\rm Tr}
\left[ \xi_L \gamma_\mu (1-\gamma_5)
       (\sqrt{Z_H} \widehat{H} - \sqrt{Z_K} \widehat{K}) \right]
\quad .
\label{weakspecial}
\end{eqnarray}
On the basis of (\ref{weakspecial})
the weak decay constant $f_{H,K}$, defined\footnote{
Our normalization is
$\langle 0|\Phi^5_v|H_v(0^-)\rangle  = \sqrt{M_H}$,
$\langle 0|\Phi_v|K_v(0^+)\rangle  = \sqrt{M_K}$.
The definition of $f_{H,K}$
corresponds to $f_\pi = \sqrt{2}F_\pi = 132$ MeV. }
through
\begin{eqnarray}
    \langle 0| \bar{q} \gamma_\mu (1-\gamma_5) Q_v|H_v(0^-)\rangle
  &=& i f_{H} M_{H} v_\mu \nonumber \quad ,\\
   \langle 0| \bar{q} \gamma_\mu (1-\gamma_5) Q_v|K_v(0^+)\rangle
  &=&  - f_{K} M_{K} v_\mu
\end{eqnarray}
is now related to the effective
coupling $G_3$ and the renormalization factors of the
heavy meson fields by
\begin{equation}
  f_{H,K} \sqrt{M_{H,K}} = \frac{\sqrt{Z_{H,K}}}{G_3}
\quad .
\label{law}
\end{equation}
We recover the familiar scaling of the weak decay constant
of heavy mesons with the heavy mass in HQET due to
heavy spin and flavor symmetry.
Note that by inserting the respective masses and $Z^i_{H,K}$--factors
we can account for flavor symmetry breaking effects and parity
splitting between the decay constants of heavy meson doublets $H,K$.

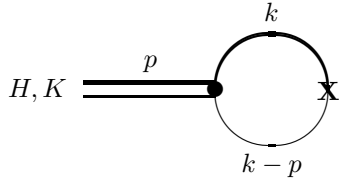
\begin{figure}[hbt]
\unitlength1cm
\begin{center}
\begin{picture}(5,3)
\thicklines
\put(0,1.6){\line(1,0){1.75}}
\put(0,1.57){\line(1,0){1.75}}
  \put(0.8,1.8){\mbox{$p$}}
\put(2.5,1.5){\oval(1.5,1.5)[t]}
\put(2.5,1.47){\oval(1.5,1.5)[t]}
  \put(2.4,2.4){\mbox{$k$}}
\thinlines
\put(2.5,1.5){\oval(1.5,1.5)[b]}
  \put(2.1,0.4){\mbox{$k-p$}}
\put(0,1.4){\line(1,0){1.75}}
\put(1.75,1.5){\circle*{0.2}}
\put(3.1,1.35){\mbox{\bf X}}
 \put(-1,1.4){\mbox{$H,K$}}
\end{picture}
\end{center}
\caption{Feynman diagram describing the
         weak decay by an arbitrary current insertion marked {\bf X}.}
\label{dia3}
\end{figure}

Due to spin symmetry of heavy quarks one can generalize
(\ref{weakspecial}) to arbitrary Dirac matrices $\Gamma$,
considering e.g. the penguin--operator
$\overline{\chi} \xi_L i\sigma_{\mu\nu} (1+\gamma_5) Q_v $
that appears
in an effective operator basis for rare $b \to s$ decays.
In our approach one can see this explicitely by calculating
the diagram in Figure \ref{dia3}.

This is completely analogous to the calculation of
the mass differences $\Delta M_{H,K}$ and leads to a
corresponding piece in the effective lagrangian
\begin{eqnarray}
  \frac{1}{2G_3} {\rm Tr} \left[ \eta^\dagger \Gamma
   (\sqrt{Z_H}\widehat{H} - \sqrt{Z_K}\widehat{K}) \right]
\end{eqnarray}
for arbitrary $\Gamma$.
The special choice
$\eta^\dagger \Gamma= \eta_L^{\dagger\,\mu} \xi_L \gamma_\mu
(1-\gamma_5)$ then indeed reproduces (\ref{weakspecial}).

In unitary gauge (\ref{unit}) we finally
arrive at the following
expressions describing electroweak decays of heavy mesons through
bosonized currents
\begin{eqnarray}
{\cal L}^{weak} &=&
\eta_L^{\mu\dagger} J_{\mu L}
+ \eta_{\mu\nu}^\dagger J^{\mu\nu}  + h.c.
\label{weda} \quad ,\\
J_{\mu L}  &=& \frac{\sqrt{M_H}f_H}{2}
{\rm Tr} \left[ \xi^\dagger \gamma_\mu (1-\gamma_5)
         \widehat{H} \right]
\nonumber \\
&&     -  \frac{\sqrt{M_K}f_K}{2}
{\rm Tr} \left[ \xi^\dagger \gamma_\mu (1-\gamma_5)
         \widehat{K} \right]
\quad ,\\
J^{\mu\nu} &=& \frac{\sqrt{M_H}f_H}{2}
{\rm Tr} \left[ \xi^\dagger i \sigma^{\mu\nu} (1+\gamma_5)
         \widehat{H} \right]
\nonumber \\
&&     -  \frac{\sqrt{M_K}f_K}{2}
{\rm Tr} \left[ \xi^\dagger i \sigma^{\mu\nu} (1+\gamma_5)
         \widehat{K} \right]
\quad .
\end{eqnarray}
Our result coincides with expressions given on the basis of
symmetry arguments in \cite{Pha}.
However a term like ${\rm Tr}
\left[ \xi^\dagger \gamma_5 H (V_\mu - {\cal V}^\pi_\mu) \right]
$ as considered sometimes in literature is absent.
Let us stress that expression (\ref{weda}) describes as well direct
decays into multi--$\pi$ states, if one expands $\xi^\dagger$ in
unitary gauge. In phenomenological applications these so--called
Callan--Treiman contributions play an important role in
weak semileptonic decays like $B \to \pi e \nu$.

\subsubsection{Isgur--Wise function}

In the heavy quark limit
the Isgur--Wise function $\xi(v\cdot v')$ describes as
a universal form factor the
matrix elements of electroweak heavy--to--heavy currents
between two heavy mesons of different velocities $H_v,H_{v'}$.
It is defined as\footnote{Recall that $Q=b,c,..$ such that
(\ref{iwdef}) describes both heavy flavor diagonal and
non--diagonal transitions.}
\begin{equation}
\langle H_{v'}(0^-,1^-)|
   \overline{Q}_{v'} \Gamma Q_v
|H_v(0^-,1^-)\rangle
   = \xi(v\cdot  v') {\rm tr_D}
  \left[ {\overline{\cal H}}_{v'} \Gamma {\cal H}_v \right]
\quad ,
\label{iwdef}
\end{equation}
where $\Gamma$ is an arbitrary Dirac matrix and ${\cal H}_v$ denotes
the matrix
representation of a heavy pseudoscalar and a heavy vector
meson with polararization vector $\varepsilon^\mu$
\begin{equation}
{\cal H}_v = \frac{1+\slash{v}}{2} \sqrt{M_H}
(i\gamma_5 + \slash{\varepsilon})
\quad .
\end{equation}
An equivalent relation holds for matrix elements between the
parity conjugate heavy mesons $K$.

\begin{figure}[hbt]
\begin{center}
\unitlength1cm
\begin{picture}(5,3)
\thicklines
\put(0,1.6){\line(1,0){1.75}}
\put(0,1.57){\line(1,0){1.75}}
  \put(0.8,1.8){\mbox{$p$}}
\put(3.25,1.6){\line(1,0){1.75}}
\put(3.25,1.57){\line(1,0){1.75}}
  \put(4.1,1.8){\mbox{$p'$}}
\put(2.5,1.5){\oval(1.5,1.5)[t]}
\put(2.5,1.47){\oval(1.5,1.5)[t]}
  \put(1.65,2.05){\mbox{$k$}}
  \put(3.15,2.05){\mbox{$k'$}}
\put(2.35,2.1){\mbox{\bf X}}
\thinlines
\put(2.5,1.5){\oval(1.5,1.5)[b]}
  \put(2.1,0.4){\mbox{$k-p$}}
\put(0,1.4){\line(1,0){1.75}}
\put(3.25,1.4){\line(1,0){1.75}}
\put(1.75,1.5){\circle*{0.2}}
\put(3.25,1.5){\circle*{0.2}}
 \put(-1,1.4){\mbox{$H_v$}}
 \put(5.1,1.4){\mbox{${\overline{H}}_{v'}$}}
\end{picture}
\end{center}
\caption{Feynman--diagram for the Isgur--Wise function.}
\label{dia4}
\end{figure}
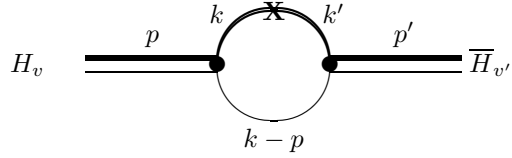

In our approach it is straightforward to
calculate the Isgur--Wise function by
simply introducing an appropriate source--term (\ref{source2})
$$
\eta^\dagger_{vv'}(\overline{Q}_{v'}  \Gamma Q_v)  + h.c.
$$
from the very beginning and
differentiating with respect to $\eta^\dagger_{vv'}$.
This gives rise to the
Feynman diagram shown in Figure \ref{dia4} and yields for the
Isgur--Wise function

\begin{eqnarray}
\xi(v\cdot v') =
Z_H
\frac{iN_c}{16\pi^4} \int^{reg}
&
d^4k&
 \frac{{\rm tr_D}
\left[(\slash{k} -\slash{p} + m) \overline{H}_{v'} \Gamma H_v\right]}{
{\rm tr_D} \left[ \overline{H}_{v'} \Gamma H_v \right]}
\nonumber \\
&
\times &
\frac{1}{((k-p)^2 - m^2) (v \cdot k + i\epsilon)
           ( v' \cdot k' + i\epsilon)}
. \nonumber \\
\end{eqnarray}

Due to spin symmetry of heavy mesons the result is indeed
independent of the particular form of the Dirac matrix $\Gamma$.
Performing the same calculational steps as for
the determination of  $Z_H$ (see section \ref{zh}) we arrive at
the following
expression for $\xi(\omega)$
in terms of the integrals $I_2,I_3,I_5(\omega)$
(see appendix \ref{appC}),
\begin{eqnarray}
\xi(\omega) = Z_{H}
         \left( \frac{2}{1+\omega} I_3  + m I_5(\omega) \right)
\quad ,
\end{eqnarray}
where the light flavor index has been dropped.
The functional dependence of the integral $I_5$ on the
momentum transfer $\omega=v\cdot v'$ is given by the
function $r(\omega)$,
\begin{equation}
I_5(\omega)= 2 I_2 r(\omega)
\quad ,
\qquad r(\omega)=\frac{\ln(\omega + \sqrt{\omega^2 -1})}{
             \sqrt{\omega^2-1}}
\end{equation}
with $I_5(1)=2 I_2$ and $I_5'(1)=-2/3 I_2$. We explicitely
get the correct normalization $\xi(\omega=1)=1$, and the slope
of the Isgur--Wise function at zero recoil is obtained as
\begin{eqnarray}
  \xi'(\omega=1) =
    Z_{H} \left( -\frac{1}{2} I_3  - \frac{1}{3} 2 m I_2 \right)
\quad .
\end{eqnarray}

The Isgur--Wise form factor for members of the $(0^+,1^+)$--multiplet,
which we refer to as $\xi_K(\omega)$,
is not related to $\xi(\omega)$ by heavy quark symmetries. Here
an analogous calculation yields
\begin{eqnarray}
\xi_K(\omega) = Z_{K}
         \left( \frac{2}{1+\omega} I_3  - m I_5(\omega) \right)
\quad .
\end{eqnarray}

It is worth mentioning that this result includes
effects of a physical cut--off $\Lambda$ (defining the scale
of chiral symmetry breaking) through the integral
$I_3$ that would be absent in renormalization schemes like
$\overline{MS}$.
A numerical discussion of the slope of
the Isgur--Wise function at the non--recoil point
$v=v'$ will be given in the following section 4.

\subsubsection{The effective lagrangian}

Finally, let us collect all obtained contributions to
the effective lagrangian for renormalized
light and heavy meson fields,
including strong interaction couplings and electroweak
currents of heavy mesons
\begin{eqnarray}
{\cal L}& =& {\cal L}^{light}
           + {\cal L}_0^{heavy}
           + {\cal L}_{V/A}^{heavy}
           + {\cal L}_{\pi}^{heavy}
\nonumber\\
&&         + (\eta_L^{\mu\dagger} J_{\mu L}
           + \eta^{\mu\nu\,\dagger} J^{\mu\nu}  + h.c.)
           + \eta^\dagger_{vv'} J_{vv'}
\quad , \\
{\cal L}^{light} & = &
  - \frac{1}{2} {\rm tr_F}
    \left[\widehat{V}_{\mu\nu}^2 \right]
  + a F_\pi^2 {\rm tr_F}
   \left[ (g_V\widehat{V}_\mu - {\cal V}_\mu^{\hat{\pi}})^2
   \right]
  \nonumber \\
  && + {\rm tr_F} \left[ \widehat{A}_{\mu\nu}^2 \right]
     + M_A^2 {\rm tr_F} \left[ (\widehat{A}_\mu)^2
				\right]
  \nonumber\\
  && + F_\pi^2 {\rm tr_F} \left[ ({\cal A}^{\hat{\pi}})^2 \right]
     + \frac{F_\pi^2 M_\pi^2}{4} {\rm tr_F}
       \left[\xi^2 + (\xi^\dagger)^2 \right]
\quad ,\\
{\cal L}_0^{heavy} & = &
  - {\rm tr_D}
      \left[ \widehat{\overline{H}}^i (iv\cdot \partial -
                \Delta M_H^i)
                   \widehat{H}^i
      \right]
\nonumber \\
&&
  + {\rm tr_D}
       \left[ \widehat{\overline{K}}^i (iv\cdot \partial -
                  \Delta M_K^i)
                   \widehat{K}^i
       \right]
\quad ,\\
{\cal L}_{V/A}^{heavy} & = &
   + \lambda^{ij}_1 g_V^{ij}
     {\rm tr_D} \left[ \widehat{\overline{H}}^j \widehat{H}^i \right]
                v \cdot \widehat{V}^{ij}
   - \lambda^{ij}_2 g_V^{ij}
     {\rm tr_D} \left[ \widehat{\overline{K}}^j \widehat{K}^i \right]
                v \cdot \widehat{V}^{ij}
\nonumber \\
&& + \lambda^{ij}_3 g_V^{ij}
     {\rm tr_D}
     \left[ \widehat{\overline{H}}^j \widehat{H}^i
            \widehat{\slash{A}}^{ij} \gamma_5 \right]
   + \lambda^{ij}_4 g_V^{ij} {\rm tr_D}
     \left[ \widehat{\overline{K}}^j \widehat{K}^i
     \widehat{\slash{A}}^{ij} \gamma_5 \right]
\nonumber \\
&& - \lambda^{ij}_5 g_V^{ij}  {\rm tr_D}
     \left[ \widehat{\overline{K}}^j \widehat{H}^i
            \widehat{\slash{V}}^{ij} \right]
   + \lambda^{ij}_6 g_V^{ij} {\rm tr_D}
     \left[ \widehat{\overline{K}}^j \widehat{H}^i
            \widehat{\slash{A}}^{ij} \gamma_5 \right] + h.c.
\quad ,\\
{\cal L}_{\pi}^{heavy} & = &
   +  g_{HHA}^{ij} {\rm tr_D}
     \left[ \widehat{\overline{H}}^j \widehat{H}^i
            \slash{{\cal A}}^{\hat{\pi}\,ij} \gamma_5 \right]
   -  g_{KKA}^{ij} {\rm tr_D}
     \left[ \widehat{\overline{K}}^j \widehat{K}^i
            \slash{{\cal A}}^{\hat{\pi}\,ij} \gamma_5 \right]
\nonumber \\
&& -  g_{HKA}^{ij} {\rm tr_D}
     \left[ \widehat{\overline{K}}^j \widehat{H}^i
            \slash{{\cal A}}^{\hat{\pi}\,ij} \gamma_5 \right]
+ h.c.
\quad ,\\
J_{\mu L}  &=& \frac{\sqrt{M_H}f_H}{2}
{\rm Tr} \left[ \xi^\dagger \gamma_\mu (1-\gamma_5)
         \widehat{H} \right]
\nonumber \\
&&     -  \frac{\sqrt{M_K}f_K}{2}
{\rm Tr} \left[ \xi^\dagger \gamma_\mu (1-\gamma_5)
         \widehat{K} \right]
\quad ,\\
J^{\mu\nu} &=& \frac{\sqrt{M_H}f_H}{2}
{\rm Tr} \left[ \xi^\dagger i \sigma^{\mu\nu} (1+\gamma_5)
         \widehat{H} \right]
\nonumber \\
&&     -  \frac{\sqrt{M_K}f_K}{2}
{\rm Tr} \left[ \xi^\dagger i \sigma^{\mu\nu} (1+\gamma_5)
         \widehat{K} \right]
\quad ,\\
J_{vv'} & = &
\xi(v\cdot  v') {\rm tr}
  \left[ \widehat{\overline{H}}_{v'} \Gamma \widehat{H}_v \right]
-\xi_K(v\cdot  v') {\rm tr}
  \left[ \widehat{\overline{K}}_{v'} \Gamma \widehat{K}_v \right]
\quad .
\end{eqnarray}

The above given effective meson lagrangian and currents, which we have
obtained from the bosonization of the NJL model, is our main formal
result. They describe the low--energy dynamics of light and heavy
mesons and in particular the interplay between spontaneous breaking of
chiral symmetry and heavy quark symmetry. Such effective meson
lagrangians have previously been written down only on phenomenological
grounds \cite{Pha}. Let us emphasize that in the meson
lagrangian and bosonized currents obtained here, the heavy meson decay
and coupling constants are all expressed in terms of the few
parameters of
the NJL model, which can be entirely fixed
from light meson data ($G_1$, $G_2$, $\widehat{m}_0$, $\Lambda$)
and heavy meson masses ($G_3$).


\section{Numerical Discussion}
\label{num}

In the previous chapters we have extended the NJL--model
of the light quark sector (determining a universal cut--off
$\Lambda=1.25$ GeV and light constituent quark masses
$\widehat{m}={\rm diag}(300,300,510)$ MeV) to the heavy
quark sector, introducing a further coupling constant
$G_3$.
This new parameter can be fixed from the experimental
mass splitting  $M_{D_s} - M_D = \Delta M_H^s -
\Delta M_H^u \approx 100 $ MeV \cite{PD}\footnote{
We use the values and notations of the Particle Data Group \cite{PD}
where $D^{(*)}$ stands for $H^{u,d}$, $D_s^{(*)}$ for $H^s$,
$D_1^{(*)}$ for $K^{u,d}$, $D_{s1}^{(*)}$ for $K^s$.
$B$ mesons are denoted analogously.}.
Varying the coupling $G_3$ in a range of
$5$ GeV$^{-2} \leq G_3 \leq 9$ GeV$^{-2}$,
we find a favoured value
$G_3 = 8.7$ GeV$^{-2}$,
which then predicts a weak decay constant $f_B = 180$ MeV
in perfect agreement with other theoretical approaches,
like lattice QCD \cite{Latt} or QCD sum rules
\cite{Sum}.
We present the results for the range
$5$ GeV$^{-2} \leq G_3 \leq 9$ GeV$^{-2}$
in table \ref{tableG3}.
\begin{table}[hbt]
\begin{center}
\begin{tabular}{|c||c|c|c|c|c|}
\hline
$G_3 [ \mbox{GeV}^{-2} ] $ &5&6&7&8&9 \\
\hline
$\Delta M_H^u \left[ \mbox{MeV}\right]$ &
750 & 540 & 390 & 280 & 200 \\
$\Delta M_H^s \left[ \mbox{MeV}\right]$ &
990 & 730 & 540 & 400 & 290 \\
\hline
$f_B \left[ \mbox{MeV}\right]$ &
310 & 260 & 220 & 190 & 170 \\
$f_{B_s} \left[ \mbox{MeV}\right]$ &
340 & 280 & 240 & 210 & 190 \\
\hline
\end{tabular}
\end{center}
\caption{Mass differences and decay constants of heavy
mesons as functions of the coupling constant $G_3$.}
\label{tableG3}
\end{table}

One sees that our fit requires a slightly larger value for $G_3$
compared to $G_1 = 5.7$ GeV$^{-2}$ obtained from the
light sector. (Recall the analogous situation in the light meson
sector where $G_2$ has to be chosen larger than $G_1$ in order to
satisfactorally describe the $\rho$--$a_1$--sector.)

Fixing $G_3 = 8.7$ GeV$^{-2}$ we can predict heavy quark
masses\footnote{
For heavy flavors we do not distinguish between current quark masses
and constituent quark masses.}
within our model using averaged values
$M_B \approx 5.3$ GeV, $M_D \approx 1.9$ GeV \cite{PD}  as input and
$\Delta M_H^u = 220$ MeV,
\begin{eqnarray}
m_b = M_B - \Delta M_H^{u}& \approx & 5.1 \mbox{ GeV}
\quad ,\\
m_c = M_D - \Delta M_H^{u}& \approx & 1.7 \mbox{ GeV}
\quad ,
\end{eqnarray}
to be compared with the respective masses derived from
$b\bar{b}$, $c\bar{c}$ bound states
\begin{eqnarray}
m_b \approx \frac{M_\Upsilon}{2} &=& 4.73 \mbox{ GeV} \nonumber
\quad ,\\
m_c \approx \frac{M_{J/\psi}}{2} &=& 1.55 \mbox{ GeV}
\quad .
\nonumber
\end{eqnarray}

\vspace{0.5cm}

In HQET weak decay constants within the
same spin--flavor multiplet scale with the heavy
meson masses
$ \sqrt{M_H} f_H = const. $,
$ \sqrt{M_K} f_K = const. $
Observables of different multiplets are, of course,
not related by heavy quark symmetries, so one expects
$ \sqrt{M_H} f_H \neq \sqrt{M_K} f_K $.
In our approach this is connected with
different renormalization factors $Z_{H,K}$
appearing in (\ref{law}).
In particular, we can account explicitly for
the dependence on different light quark masses.
The experimental determination of $f_D, f_B$ is
still vague.
Lattice--calculations \cite{Latt} and QCD sum rules
\cite{Sum} give values around
$ f_D \approx  200 $ MeV and $ f_B \approx 180$ MeV.
Both approaches  predict large $\Lambda/m_c$
corrections to the HQET scaling law when including  effects
of non--leading operators in the HQET lagrangian (\ref{L-HQET}).
Nevertheless in the ratio
\begin{eqnarray}
R=\frac{f_{H}^s}{f_{H}^u}
\end{eqnarray}
which accounts for light $SU(3)_F$ symmetry breaking
effects, these corrections are expected to
cancel and our result for this ratio is independent of $G_3$,
$R \approx 1.1$.
Other approaches based on one--loop calculations in
chiral perturbation theory \cite{GrWi}, lattice
simulations \cite{Latt} and QCD sum rules \cite{Sum}
yield  similar
results $R=1.1 $ -- $ 1.2$.

Note that usual hard gluon QCD--corrections
can be taken into account by
the scale dependence of the strong coupling constant
computed within a leading log approximation in HQET
\cite{HQET}.
In our approach the required scale
matching should be done at the scale $\Lambda$,
\begin{eqnarray}
\sqrt{M_B} f_B
&=&  \left( \frac{\alpha_s(\Lambda)}{\alpha_s(M_B)}\right)^{-6/25}
       \sqrt{M_H} f_H
\quad , \nonumber
 \\
\sqrt{M_D} f_D
& =&  \left( \frac{\alpha_s(\Lambda)}{\alpha_s(M_D)}\right)^{-6/25}
       \sqrt{M_H} f_H
\quad , \nonumber
\end{eqnarray}
to give effects up to 10--15\%.

Concerning the heavier $(0^+,1^+)$ states we observe that
$Z_K$ in (\ref{Zfac}) together with $\Delta M_K$ in
(\ref{selfcon}) get unrealistically large
($\Delta M_K^u = 2050$ MeV, $\Delta M_K^s = 6120$ MeV).
We expect here essential numerical improvement from on--shell
corrections at $v\cdot p = \Delta M_K$ which are not
included in the pure gradient expansion around $v \cdot p = 0$.

\vspace{0.5cm}

Let us next focus on the Isgur--Wise function $\xi(\omega)$.
The slope parameter at the non--recoil point
is defined through
$$ \rho^2 = - \xi'(1) \quad .$$
The predictions of our model are
\begin{eqnarray}
  \xi^{u'}(\omega=1)& = &- 0.44   \to \rho = 0.67
\quad ,\\
  \xi^{s'}(\omega=1)& = &- 0.43   \to \rho = 0.66
\quad .
\end{eqnarray}
Recent fits on ARGUS and CLEO data
prefer a value of
$\rho=1.14\pm 0.23$ \cite{Nb} which is in consistency
with QCD sum rule estimates of $\rho = 1$ \cite{Sum}.

Notice that
naively calculating a triangle quark diagram in a
renormalization scheme like $\overline{MS}$ would
give $\xi(\omega)=r(\omega)$ together with a lower value
$\rho = 1/\sqrt{3} = 0.58$.
In contrast, our calculation with composite mesons uses
a physical
cut--off (related to the scale of
chiral symmetry breaking).
Following the discussion in connection with
analytic bounds on the Isgur--Wise function \cite{RT},
one would expect an increase in  $\rho$ if  additional
effects of bound states consisting of two heavy
quarks would be taken into account.

\vspace{0.5cm}

Concerning the several coupling parameters between
heavy mesons and light vector or axial vector fields,
we focus on the results for $\lambda_1^{ij}$ and
$\lambda_3^{ij}$.
The values of $\lambda_1^{ii}$ are fixed by $SU(3)_V$
symmetry to $\lambda_1^{uu} = \lambda_1^{ss} = 1$ such
that the light vector mesons couple through covariant
derivatives.
Moreover, explicit $SU(3)_F$ symmetry breaking leads to
a slight decrease in the parameter $\lambda_1^{us}$,
\begin{eqnarray}
\lambda_1^{us} &=& 0.98
\quad .
\end{eqnarray}
Let us next consider the value for $\lambda_3^{ij}$ which
is connected to the pseudoscalar meson coupling $g_{HHA}^{ij}$ in
(\ref{gHHA}) and enters into the rate for decays like
$B \to  \pi e \nu$, $D \to K e \nu$,
\begin{eqnarray}
\lambda_3^{uu} &=& - 0.33 \quad ,\quad g_{HHA}^{uu} = - 0.17
  \quad ,\\
\lambda_3^{us} &=& - 0.45 \quad ,\quad g_{HHA}^{us} = - 0.22
  \quad ,\\
\lambda_3^{ss} &=& - 0.59 \quad ,\quad g_{HHA}^{ss} = - 0.29
  \quad .
\end{eqnarray}
{}From the process
$D^{*+} \to D^0 \pi^+$, there exists an
upper bound for $g_A = g_{HHA}^{uu}$
\begin{eqnarray}
&& \Gamma(D^{*+} \to D^0 \pi^+) =
\frac{g_A^2 |\vec{p}_\pi|^3}{6 \pi F_\pi^2} < 0.131
  \mbox{ MeV \cite{AC}} \\
&& \mbox{leading to} \quad  g_A^2 < 0.5 \quad,
\end{eqnarray}
which is respected by our result.
Recent analysises of CLEO data for $D^*$ branching fractions
including electromagnetic interactions using a
phe\-no\-me\-no\-lo\-gi\-cal
heavy meson chiral
lagrangian \cite{gA} find a larger value of $g_A$
given by $g_A^2=0.34 \pm 0.48$ including still a large error.


\section{Conclusions}

In this work we have studied the properties of composite
heavy mesons within the framework of an extended QCD--motivated
NJL--model.
The light quark flavor dynamics of this model is governed
by chiral $SU(3)_L \times SU(3)_R$ symmetry of QCD
and its spontaneous and explicit breaking, while the heavy quark
sector incorporates the new heavy quark symmetries.

By applying path integral bosonization techniques and
performing a gradient expansion, we have derived an
effective low--energy lagrangian of light and heavy mesons.

Except for the heavy--light quark interaction strength $G_3$ we
have adjusted the parameters of our model (light quark masses,
interaction constants $G_1$, $G_2$ and the universal cut--off)
by fitting light meson properties.
For the $(0^-,1^-)$ spin--flavour multiplet, heavy meson
masses and weak decay constants are then successfully described
with a heavy--light
four--quark interaction constant
$G_3 = 8.7 $ GeV$^{-2}$.

The coupling of
heavy $(0^-,1^-)$ mesons to the axial current of Goldstone bosons
is calculated as $g_A=-0.17$, lying within experimental bounds
$g_A^2 < 0.5$, while light vector mesons couple by covariant
derivatives.

Next, by bosonization of electroweak currents of heavy mesons,
the weak decay constants have been determined as
$f_B = 180$ MeV, $f_D = 300$ MeV. While
the value for $f_B$ coincides with recent lattice estimates,
the value for $f_D$ is somewhat large and
expected to be improved by including
$1/m_c$ corrections.
The effect of explicit $SU(3)_F$ breaking is estimated through
the ratio $R=f_H^s/f_H \approx 1.1$ in
consistency with other predictions.

Finally, the Isgur--Wise function is calculated with a
slope parameter of $\rho \approx 0.67$
which is somewhat smaller than experimental findings.

Although conceptionally analogous to the $(0^-,1^-)$
case, our numerical results for the heavy $(0^+,1^+)$
spin--flavour multiplet are less satisfactory.
Work on a possible improvement by performing an on--shell
calculation expanding around $v \cdot p = \Delta M$
(instead of $v \cdot p = 0$ ) is in progress.

In summary we have shown that the synthesis of chiral
and heavy spin--flavor symmetries within an extended
NJL model leads to an effective lagrangian offering a
unified description of strong and weak interactions
of heavy and light mesons.
The above bosonization approach could be
further generalized into two directions:
First, it would be interesting to study non--local
NJL--interactions based on non--perturbative
gluon propagators by using the bilocal field
approach (see for example \cite{Eb94}).
Secondly, to get a complete picture of hadron dynamics,
baryons should be included as well.

\section*{Note added}

During completion of this work investigations of
Bardeen and Hill \cite{BH93}
and No\-wak, Rho and Zahed \cite{No93},
partially studying similar questions,
were brought to our attention.
Our work differs, however, in the used model, in important
con\-ceptual
aspects and naturally also in the numerical results.

\section*{Acknowledgement}
It is a pleasure to thank Thomas Mannel for enjoyable discussions
on HQET.
One of us (T.F.) would like to thank the HEP--group of
TH Darmstadt for useful conversations during his visit in winter
1993.



\appendix
\renewcommand{\theequation}{\Alph{section}.\arabic{equation}}
\setcounter{equation}{0}

\section{Basic Formulae of HQET}
\label{appA}

Following \cite{HQET}, we study heavy quarks in the
full QCD lagrangian given by
$$ {\cal L} =  \overline{Q} (i\slash{D} - m_Q) Q \quad ,     $$
where the
covariant derivative of QCD reads
$ D_\mu = \partial_\mu + i g A^\alpha_\mu \lambda_c^\alpha/2 $.

It is convenient to introduce
upper ($\varphi$) and lower components ($\vartheta$)
of the heavy quark spinor $Q$
\begin{eqnarray}
\varphi & = & P_+ Q
\quad ,\\
\vartheta & = & P_- Q
\end{eqnarray}
with
\begin{eqnarray}
\slash{v} \varphi &=&  \varphi
\quad ,\\
\slash{v} \vartheta &=& -\vartheta
\quad ,
\end{eqnarray}
where $P_\pm$ are projectors on the heavy quark velocity
$v^\mu$ ($v^2=1$),
$$ P_\pm = \frac{1 \pm \slash{v}}{2}\quad , $$
satisfying the following relations
\begin{eqnarray}
P_+ \gamma_\mu P_+ &=& P_+ v_\mu \quad , \\
P_- \gamma_\mu P_- &=& -P_- v_\mu \quad , \\
P_+ \gamma_\mu P_- &=& P_+ (\gamma_\mu - v_\mu) \quad , \\
P_- \gamma_\mu P_+ &=& P_- (\gamma_\mu + v_\mu)
\quad .
\end{eqnarray}

It is further convenient to define a longitudinal and a transversal
part of the covariant derivative
\begin{eqnarray}
   \slash{D}& =& \slash{v}(v \cdot D) + \slash{D}^\perp \quad , \\
  \slash{D}^\perp & =&
	\gamma^\mu (g_{\mu\nu} - v_\mu v_\nu ) D^\nu \quad , \\
  \{\slash{D}^\perp , \slash{v} \}& = &0
\quad .
\end{eqnarray}

One now has to chose between two kinds of parametrizations,
describing particles or anti--particles
\begin{eqnarray}
  \mbox{particle: } && \varphi= e^{-im_Q (vx)} h^+_v
                \qquad \vartheta = e^{-im_Q (vx)}   H^+_v \quad , \\
  \mbox{antiparticle: } && \varphi= e^{+im_Q (vx)} H^-_v
                    \qquad \vartheta= e^{+im_Q (vx)} h^-_v
\quad .
\end{eqnarray}
(The relevant degree of freedom for particles $h^+_v$ is denoted
as $Q_v$ in the text.)
The lagrangian in the particle sector then reads
\begin{eqnarray}
  {\cal L} &=&
             \bar{h}^+_v i v\cdot D h^+_v
            -\bar{H}^+_v (i v\cdot D + 2 m_Q) H^+_v \nonumber \\
      &&    +\bar{h}^+_v i \slash{D}^\perp  H^+_v
            +\bar{H}^+_v i \slash{D}^\perp  h^+_v
\quad ,
\end{eqnarray}
where the field $H^+_v$ carries twice the heavy quark mass.
The integration over $H^+_v$ in the concerning
generating functional of QCD can easily be carried out,
 defining an effective lagrangian
\begin{eqnarray}
{\cal L}^{HQET} &=& \bar{h}_v^+ i (v\cdot D) h_v^+
                 +\bar{h}_v^+ i \slash{D}^\perp
                          \frac{1}{i(v\cdot D) + 2m_Q - i \epsilon}
                              i \slash{D}^\perp h_v^+ \\
&=& \bar{h}_v^+ i(v\cdot D)h_v^+
   + {\cal K}_v +
     {\cal M}_v + O(m_Q^{-2})
\quad ,
\end{eqnarray}
where the two operators
${\cal K}_v$, ${\cal M}_v$ of order $m_Q^{-1}$
\begin{eqnarray*}
 {\cal K}_v & = &
  - \frac{1}{2m_Q} \bar{h}_v^+ D^\mu (g_{\mu\nu} - v_\mu v_\nu)
  D^\nu h_v^+ \quad , \\
{\cal M}_v &=&
  - \frac{g}{4m_Q} \bar{h}_v^+
  \sigma^{\mu\nu}F_{\mu\nu} h_v^+
\end{eqnarray*}
can be identified in
the rest--frame $\vec{v} = 0$ with the non--relativistic kinetic
energy and the chromomagnetic Pauli--term, repectively
($F^{\mu\nu}$ denotes
the gluonic field strength tensor).

The $SU(2)$ spin--symmetry, that arises in the limit
$m_Q \to \infty$, can
be implemented by generalizing the Pauli matrices $\sigma_i$ in terms
of  velocity--dependent generators
$$ S_i(v) = \gamma_5 \slash{v} \slash{\varepsilon}_i $$
with $\varepsilon_i$ being
suitable polarization vectors satisfying
$v\cdot \varepsilon_i = 0$ and $\varepsilon_i^2 = -1$.
Their effect on heavy
quark spinors $u_{\uparrow,\downarrow}$ reads
\begin{eqnarray}
    S_3(v) u_\uparrow (v) = u_\uparrow (v) &\quad , \qquad&
    S_3(v) u_\downarrow(v)= - u_\downarrow (v)
\end{eqnarray}

\setcounter{equation}{0}
\section{Fierz Transformations}

There exist two types of Fierz--rearrangements of the
color matrices in the current--current interaction
(\ref{int}), defined by \cite{Eb94}
\begin{eqnarray}
\sum_{\alpha = 1}^{N_c^2 -1}
\frac{(\lambda^\alpha_c)_{ij}}{2}
\frac{(\lambda^\alpha_c)_{kl}}{2}
&=&
\frac{1}{2} \left( 1 - \frac{1}{N_c^2} \right) \delta_{il} \delta_{kj}
-\frac{1}{N_c} \sum_{\alpha=1}^{N_c^2 - 1}
\frac{(\lambda^\alpha_c)_{il}}{2}
\frac{(\lambda^\alpha_c)_{kj}}{2}
\label{fir} ,
\\
\sum_{\alpha = 1}^{N_c^2 -1}
\frac{(\lambda^\alpha_c)_{ij}}{2}
\frac{(\lambda^\alpha_c)_{kl}}{2}
&=&
\frac{1}{2} \left( 1 - \frac{1}{N_c} \right) \delta_{il} \delta_{kj}
+\frac{1}{2N_c} \epsilon_{mik} \epsilon_{mlj}
\label{sec}
\quad .
\end{eqnarray}
The first identity transforms the current--current interaction
into an attractive color singlet $(q\bar{q})_{1c}$ interaction
and a repulsive
color $(N_c^2-1)$--plet interaction.
For later applications with diquarks within composite baryons,
we prefer the second alternative (\ref{sec}), where besides of
an attractive interaction in the $(q\bar{q})_{1c}$ channel, the
second part yields attraction in the color--antisymmetric
$(qq)$ channel.
In the limit $N_c \to \infty$ the color singlet terms in (\ref{fir}),
(\ref{sec}) dominate and have equal weight factors.

In the Fierz--transformation of light flavors use has also been made
of the completeness relation
\begin{equation}
\frac{1}{2} \delta_{ij} \delta_{kl} =
\sum_{a=0}^{N_F^2 - 1}
\frac{(\lambda^a_F)_{il}}{2} \frac{(\lambda^a_F)_{kj}}{2}
\quad ;
\qquad (\frac{\lambda^0_F}{2}=\sqrt{\frac{1}{2 N_F}} {\, \bf 1}_F)
\quad.
\end{equation}

Finally, the Fierz--transformations of Dirac matrices read
\begin{eqnarray}
(\gamma^\mu)_{ij} (\gamma_\mu)_{kl}
&=& \sum_{\alpha=1}^4
     (\hat{O}^\alpha)_{il} (\hat{O}_\alpha)_{kj} \\
&=& \sum_{\alpha=1}^4
     (\hat{O}^\alpha C)_{ik} (C \hat{O}_\alpha)_{lj}
 \quad ,
\end{eqnarray}
where
\begin{equation}
\hat{O}^\alpha = \left\{ 1, i\gamma_5,
                        \frac{i}{\sqrt{2}} \gamma^\mu
			\frac{i}{\sqrt{2}} \gamma^\mu \gamma_5
		 \right\} \quad , \qquad (\alpha=1,\ldots,4)
\end{equation}
and $C=i\gamma^2\gamma^0$ is the charge conjugation matrix.

\setcounter{equation}{0}
\section{Feynman Integrals }
\label{appC}

We present here the values of several integrals needed when
calculating the functional determinant (\ref{det}).
In our approach the denominator in Euclidean
space of the light quark
propagator is regularized with proper--time methods
\begin{eqnarray}
  \frac{1}{k^2 + m^2} &=& \int_0^\infty ds e^{-(k^2 + m^2)s}
  \to \int_{1/\Lambda^2}^\infty ds e^{-(k^2 + m^2)s}
\quad ,\\
  \frac{1}{(k^2 + m^2)^2} &=& \int_0^\infty ds \, s e^{-(k^2 + m^2)s}
  \to \int_{1/\Lambda^2}^\infty ds \, se^{-(k^2 + m^2)s}
\quad ,
\end{eqnarray}
whereas the heavy quark propagator is unaffected from
regularizations\footnote{Note that in going to Euclidean space
the $i\epsilon$--prescription in the heavy quark propagator
must be treated properly.}.
One obtains
\begin{eqnarray}
I_1^i &=& \frac{iN_c}{16\pi^4} \int^{reg}
          \frac{d^4k}{k^2 - (m^i)^2} \nonumber \\
    &=& \frac{N_c}{16 \pi^4}
           \int_{1/\Lambda^2} ds \int d^4k e^{-(k^2+(m^i)^2)s}
\nonumber \\
    &=& \frac{N_c}{16\pi^2} (m^i)^2 \Gamma(-1,(m^i)^2/\Lambda^2)
\quad ,\\
I_2^{ij} &=& - \frac{iN_c}{16\pi^4}
        \int^{reg} \frac{d^4k}{(k^2 - m_i^2)^(k^2-m_j^2)} \nonumber \\
    &=& \frac{N_c}{16\pi^4}
    \int_0^1 dx \int_{1/\Lambda^2} ds \, s
      \int d^4k e^{-(k^2+xm_i^2+(1-x)m_j^2)s} \nonumber \\
    &=& \frac{N_c}{16\pi^2}
        \int_0^1 dx \Gamma(0,(xm_i^2+(1-x)m_j^2)/\Lambda^2)  \quad ,\\
I_3^i &=& - \frac{iN_c}{16\pi^4} \int^{reg}
            \frac{d^4k}{(k^2 - (m^i)^2)(v\cdot k + i\epsilon)}
\nonumber \\
    &=& \frac{iN_c}{16\pi^4}
    \int_{1/\Lambda^2} ds \int d^4k e^{-(k^2+(m^i)^2)s}
                               \frac{1}{k_4 + i\epsilon} \nonumber \\
    &=& \frac{N_c}{16\pi^3}
    \int_{1/\Lambda^2} ds \int d^3k e^{-(k^2+(m^i)^2)s}
                                \nonumber \\
    &=& \frac{N_c}{16\pi^2} \sqrt{\pi} m
	\Gamma(-1/2,(m^i)^2/\Lambda^2)
         \quad ,\\
I_4^{ij} &=&  \frac{iN_c}{16\pi^4} \int^{reg}
            \frac{d^4k}{(k^2 - (m^i)^2)(k^2 - (m^j)^2)
	(v\cdot k + i\epsilon)}
 \nonumber \\
         &=& \frac{iN_c}{16\pi^4} \int_0^1 dx
    \int_{1/\Lambda^2} ds \, s \int d^4 k \,
	         e^{-(k^2+xm_i^2+(1-x)m_j^2)s}
                 \frac{1}{k_4 + i\epsilon} \nonumber \\
         &=& \frac{N_c}{16\pi^3} \int_0^1 dx
    \int_{1/\Lambda^2} ds \, s \int d^3k  \,
             e^{-(k^2+xm_i^2+(1-x)m_j^2)s}
 \nonumber \\
         &=& \frac{N_c}{16\pi^2} \sqrt{\pi} \int_0^1 dx
             \frac{\Gamma(1/2,(xm_i^2+(1-x)m_j^2)/\Lambda^2)}{
                   \sqrt{xm_i^2+(1-x)m_j^2}}
         \quad ,\\
I_5(\omega=v\cdot v') &=& \frac{iN_c}{16\pi^4} \int^{reg}
        \frac{d^4k}{(k^2-m^2)(v\cdot k + i\epsilon)
	(v'\cdot k + i\epsilon)}
 \nonumber \\
 & = &\frac{-N_c}{16\pi^4}
        \int_{1/\Lambda^2} ds \int d^4k e^{-(k^2+m^2)s}
        \frac{1}{(v\cdot k + i\epsilon)
                 (v'\cdot k + i\epsilon)}
 \nonumber \\
    &=& \frac{-N_c}{16\pi^4}
        \int_{1/\Lambda^2} ds \int d^4k e^{-(k^2+m^2)s}
        \frac{1}{(e\cdot k + i\epsilon)^2} \nonumber \\
    && \times \int_0^1 \frac{dx}{2x^2(1-\omega)+2x(\omega-1)+1}
\quad ,\qquad e^2=1
 \nonumber \\
    &=& 2 I_2 \frac{\ln(\omega+\sqrt{\omega^2-1})}{\sqrt{\omega^2-1}}
\quad .
\end{eqnarray}

Here
$$\Gamma(\alpha,x) = \int_x^\infty dt e^{-t} t^{\alpha-1}$$
denotes the incomplete gamma function.

\setcounter{equation}{0}
\section{Coupling Parameters $\lambda$}
\label{appD}
In this appendix we present the expressions for the several couplings
between two heavy and one light meson as they were
defined in (\ref{lambda}). The integrals $I_2^{ij},I_3^i,I_4^{ij}$
are found in app.\ \ref{appC}.
\begin{eqnarray}
  \lambda_1^{ij} & = &
  \sqrt{Z_H^i Z_H^j} \left(
    \frac{1}{2} ( (Z_H^i)^{-1} + (Z_H^j)^{-1} )
   -\frac{1}{2} (m^i - m^j)^2 I_4^{ij} \right)
\quad , \\
\nonumber \\
  \lambda_2^{ij} & = &
  \sqrt{Z_K^i Z_K^j} \left(
    \frac{1}{2} ( (Z_K^i)^{-1} + (Z_K^j)^{-1} )
   -\frac{1}{2} (m^i - m^j)^2 I_4^{ij} \right)
\quad , \\
\nonumber \\
  \lambda_3^{ij} & = &
  \sqrt{Z_H^i Z_H^j} \left(
    \frac{1}{6} ( I_3^i + I_3 ^j ) - ( m^i + m^j) I_2^{ij}
\right. \nonumber \\
&&
\left.
   - ( \frac{4}{3} m^i m^j + \frac{1}{6} (m_i - m_j)^2 ) I_4^{ij}
\right)
\quad , \\
\nonumber \\
\lambda_4^{ij} & = &
  \sqrt{Z_K^i Z_K^j} \left(
    \frac{1}{6} ( I_3^i + I_3 ^j ) + ( m^i + m^j) I_2^{ij}
\right. \nonumber \\
&&
\left.
   - ( \frac{4}{3} m^i m^j + \frac{1}{6} (m_i - m_j)^2 ) I_4^{ij}
\right)
\quad , \\
\nonumber \\
\lambda_5^{ij} & = &
  \sqrt{Z_H^i Z_K^j} \left(
    \frac{1}{6} ( I_3^i + I_3 ^j ) - ( m^i - m^j) I_2^{ij}
\right. \nonumber \\
&&
\left.
 + ( \frac{2}{3} m^i m^j - \frac{1}{6} (m_i - m_j)^2 ) I_4^{ij}
\right)
\quad , \\
\nonumber \\
\lambda_6^{ij} & = &
  \sqrt{Z_H^i Z_K^j} \left(
    \frac{1}{2} ( (Z_H^i)^{-1} + (Z_K^j)^{-1} )
   -\frac{1}{2} (m_i + m_j)^2 I_4^{ij} \right)
\quad .
\end{eqnarray}

\end{document}